\documentclass[sigconf]{acmart}

\AtBeginDocument{%
  }

\copyrightyear{2025} 
\acmYear{2025} 
\setcopyright{cc}
\setcctype{by}
\acmConference[CHI '25]{CHI Conference on Human Factors in Computing Systems}{April 26-May 1, 2025}{Yokohama, Japan}
\acmBooktitle{CHI Conference on Human Factors in Computing Systems (CHI '25), April 26-May 1, 2025, Yokohama, Japan}\acmDOI{10.1145/3706598.3713583}
\acmISBN{979-8-4007-1394-1/25/04}

\acmConference[CHI '25]{Make sure to enter the correct
  conference title from your rights confirmation emai}{April 26-May 1, 2025,
  2025}{Yokohama, Japan}
\acmISBN{978-1-4503-XXXX-X/18/06}

\usepackage{subcaption}
\usepackage{wrapfig}
\usepackage{adjustbox}

\begin{document}

\title{The Many Tendrils of the Octopus Map}




\author{Eduardo Puerta}
\authornote{These authors contributed equally to this work.}
\email{puerta.e@northeastern.edu}
\orcid{0000-0003-4664-4365}
\author{Shani Spivak}
\authornotemark[1]
\email{spivak.s@northeastern.edu}
\orcid{0000-0003-4664-4365}
\affiliation{
    \institution{Northeastern University}
    \city{Boston}
    \state{MA}
    \country{USA}
}

\author{Michael Correll}
\email{m.correll@northeastern.edu}
\orcid{0000-0001-7902-3907}
\affiliation{%
    \institution{Roux Institute, Northeastern University}
    \city{Portland}
    \state{ME}
    \country{USA}
}

\renewcommand{\shortauthors}{Puerta \& Spivak et al.}

\begin{abstract}
Conspiratorial thinking can connect many distinct or distant ills to a central cause. This belief has visual form in the \textit{octopus map}: a map where a central force (for instance a nation, an ideology, or an ethnicity) is depicted as a literal or figurative octopus, with extending tendrils. In this paper, we explore how octopus maps function as visual arguments through an analysis of historical examples as well as a through a crowd-sourced study on how the underlying data and the use of visual metaphors contribute to specific negative or conspiratorial interpretations. We find that many features of the data or visual style can lead to ``octopus-like'' thinking in visualizations, even without the use of an explicit octopus motif. We conclude with a call for a deeper analysis of visual rhetoric, and an acknowledgment of the potential for the design of data visualizations to contribute to harmful or conspiratorial thinking.
\end{abstract}



\begin{teaserfigure}
  \centering
  \includegraphics[width=\linewidth, alt={A world map with a giant red octopus grasping at the last 10 cities where CHI has taken place}]{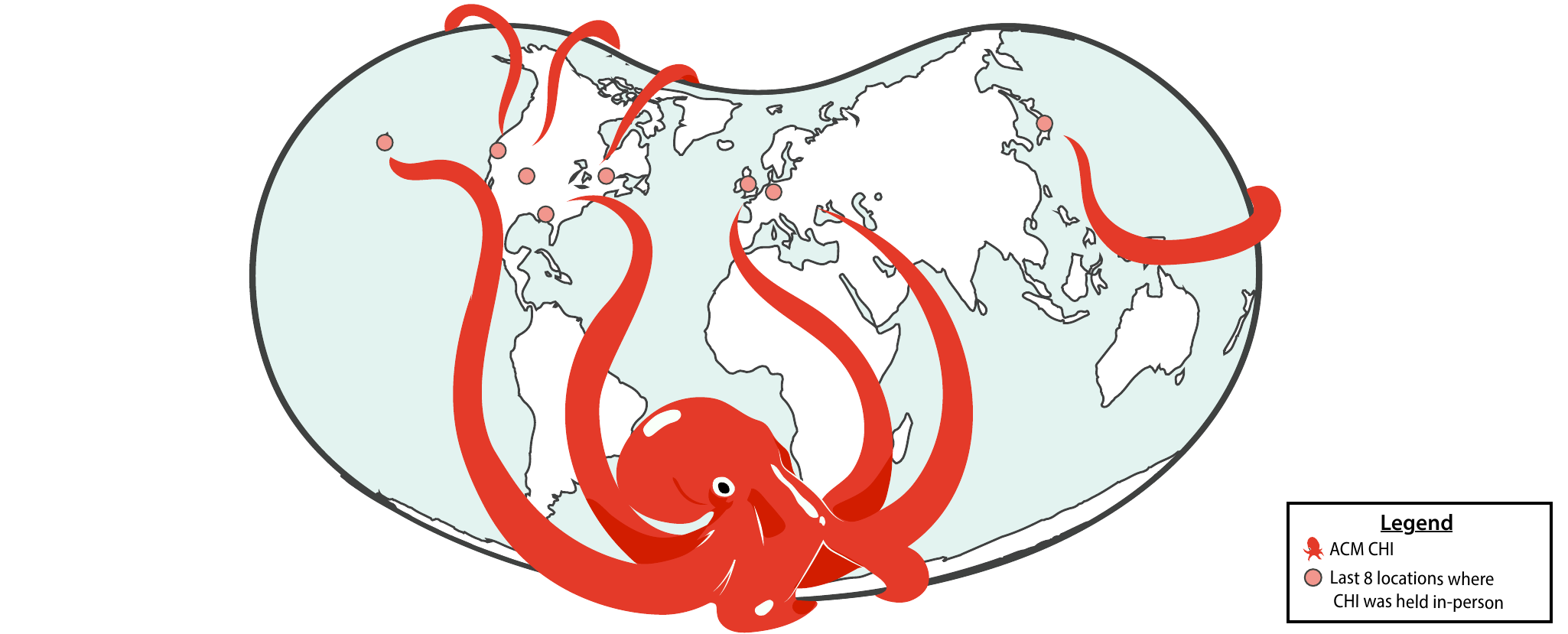}
  \Description{A world map with a giant red octopus grasping at the last eight cities where CHI has taken place}
  \caption{%
  	An octopus map of past ACM CHI conference locations. How does the visual motif of the octopus combined with the global expanse of the data points impact the perceived intent or motives of CHI?}
\label{fig:teaser}
\end{teaserfigure}

\received{20 February 2007}
\received[revised]{12 March 2009}
\received[accepted]{5 June 2009}


\maketitle

\section{Introduction}

Ironically, once you've been exposed to this visual genre, you start seeing it everywhere. An enemy, hitherto imagined as a distant or diffuse threat, is revealed as a sinister octopus, a  ``soft alien intelligence'' (De Luca Comandini~\cite{comandini1988octopus}, as quoted in Ronnberg~\cite{ronnberg2010book}) that is grasping and acquisitive. This enemy, Them (with a capital ``T''), may have a central body somewhere, with a single goal, but They also have many tentacles that extend outwards encompassing every corner of the map. Severing one tentacle may be straightforward, but such an effort is pointless without defeating the central cephalopod body. When employed in a cartographic context, we term this conspiratorial and adversarial metaphor an \textbf{octopus map} (\autoref{fig:teaser}). The octopus map, either in the literal form of an octopus menacing a geographic region, or more indirectly as a form of ``sensationalized''~\cite{muehlenhaus2012if} cartography with a central but ubiquitous force, has centuries of precedent in political cartoons and propaganda.

In this paper, we examine octopus maps through both an analysis of historical examples as well as an empirical study with the goal of assessing the visual and structural forms of these maps and their ensuing rhetorical function. We have encountered a rich space of octopus maps throughout history, from varied uses in iconography in political cartoons to more ``data-driven'' (but still ideological) cartographic forms. We find the octopuses of these maps portraying a diverse set of enemies including religious groups, ideologies, wartime adversaries, and abstract concepts. We assert that the octopus map forms a ``visual argument''~\cite{blair2012possibility} that communicates a specific view of an adversary as a grasping and acquisitive threat. We hold that this visual argument rests on two pillars: first, on the \textit{structure of the data}: that is, that certain networks and connections lead to their contents as being perceived as more intrinsically ``octopodal'' based on their graph-theoretic properties like centrality and connectedness. Second, on the \textit{visual metaphors}~\cite{ziemkiewicz2009preconceptions,ziemkiewicz2010impliedDynamics} and design choices made in representing these data, from representations intended to be neutral to intentionally sensationalized forms. In our crowdsourced evaluation, we find that, even without the use of a literal octopus motif, the various data or visual properties associated with octopus maps can nonetheless encourage threatening views of the data they represent. That is, implicit ``octopus-like'' design choices in maps can perform similar rhetorical work, and result in similar negative judgments about their subjects, as in maps with more intentionally biased or sensationalized designs.

Therefore, while the octopus maps we discuss may reflect particularly egregious or overtly biased examples of visual rhetoric, we contend that the structure of data in collaboration with visual design contribute to an intended visual argument in \textit{all} visualizations, not just a narrow slice of intentionally propagandized maps. Just as statistics can be cast as a form of ``principled argument''~\cite{abelson1995statistics}, so too can visualizations perform a great deal of rhetorical and argumentative work~\cite{hullman2011rhetoric,kennedy2016work,kostelnick2003shaping}, including emotional appeals~\cite{campbell2019feeling}. In the midst of the ``rapid rise''~\cite{kostelnick2016re} of emotional appeals in data visualization, and the ubiquity of data visualizations among conspiratorial groups~\cite{hannah2021conspiracy,lee2021viral}, we point to a need to examine the unique persuasive power of charts and maps, which often take advantage of a (falsely) assumed trustworthiness or objectivity of data~\cite{correll2019ethical,drucker2012humanistic,mode2017maps,monmonier1995draw, monmonier1996how}. Visualizations are not just the dispassionate communication of insights in data, but can also be exploited to fan the flames of fear, hatred, and xenophobia~\cite{van2020migration}.

\textit{Content warning: this paper contains images depicting racist, antisemitic, xenophobic, fascist, and other hate-based or dehumanizing ideologies and attitudes.}


\subsection{Contributions}
In this paper, we provide the following contributions:
\begin{enumerate}
    \item We provide a historical and visual overview of the octopus map as a genre of persuasive cartography and data visualization, with a particular focus on how both \textit{visual properties} and \textit{data properties} of these maps can produce ``octopus-like'' rhetorical impacts.
    \item  We deconstruct octopus maps based on these visual and structural components and conduct an empirical study to measure the impact of these properties on rhetorical outcomes.
    \item We reflect on design implications for visualization designers in light of the moral, ethical, and rhetorical concerns that arise from the explicit or implicit use of persuasive design techniques in data visualization and cartography generally.
\end{enumerate}



\section{Related Work}
\label{sec:related}

Our analysis of octopus maps draws on several existing, and occasionally interconnected, areas of research:
\begin{enumerate}
    \item Persuasive cartography (as part of larger projects on \textit{critical cartography}~\cite{crampton2018introduction}).
    \item Visual metaphors and the implicit or explicit visual rhetoric of these metaphors, with a particular focus on the visual metaphors of graph visualization, and the perception and visualization of structures in networks.
\end{enumerate}


\subsection{Persuasive Cartography}
Persuasive cartography focuses on the inherent rhetorical power of maps. That is, considering maps not as purely objective stores of equally objective data, but as intrinsically persuasive artifacts that contain implicit or explicit perspectives, biases, and interpretations~\cite{wells1992the,monmonier1995draw,monmonier1996how}. While all maps have persuasive elements, Judith Tyner defined ``persuasive maps'' as those which are created by manipulating various cartographic elements through distortion, selection, symbolization, and choice of text and title~\cite{tyner1982persuasive}. Of particular relevance to our work is prior work by Muehlenhaus on persuasive maps~\cite{muehlenhaus2012beyond,muehlenhaus2012if,muehlenhaus2013design,muehlenhaus2014going}.  Specifically, prior work on how certain visual features (like map projection, choice of symbols, and perceived dynamism) are associated with persuasive maps~\cite{muehlenhaus2013design}, and the existence of coherent visual genres of ``sensationalist''~\cite{muehlenhaus2014going} persuasive maps. In his study of 256 persuasive maps, Muehlenhaus found that while sensationalized or confrontational titles were not only rarely used, more subtle design choices, such as hue and the use of flow arrows, were common \cite{muehlenhaus2013design}.

One potential function of these sensationalist maps is to present an enemy as a threat or merely a threatening presence, which can be accomplished through the use of tools like flow lines and arrows, color, shading, and projection~\cite{monmonier1996how,muehlenhaus2013design}. These techniques can produce unintended rhetorical impacts. For instance, in work influential to our decomposition of ``octopus-like'' elements, van Houtum \& Lacy~\cite{van2020migration} refer to the ``trap'' into which many cartographers fall when portraying migration data, where techniques common in persuasive cartography xenophobically portray migrants as exterior threats that are invading or subverting a country. Even a map made by scholarly cartographers, without sufficient care, ``peddles a crude distortion of undocumented migration that smoothly splices into the xenophobic tradition of propaganda cartography''~\cite{van2020migration}. For instance, the use of large colorful arrows to represent flows of migrants into a country mirrors maps used during wartime to depict invasions by exterior foes. 


\subsection{Visual Metaphors}


At its core, and stripped of emotional appeals, the octopus map is merely a visualization of a network structure superimposed with cartography: a central node with multiple edges fanning out across the map. 
Much of the intended persuasive work therefore relies on the visual metaphors~\cite{pokojna2024language}, rhetorical framing~\cite{hullman2011rhetoric}, and conventions~\cite{kennedy2016work} used in the design and presentation of these data. We focus on prior work with respect to two notable components: the use of tentacles or tentacle-like representations of the graph structure (the \textit{structural} component), and the choice of an octopus to represent the central node (the \textit{pictorial} component).
This structure also follows the semantic visual metaphor design process described by Cruz, consisting of the ``adaption of the structural metaphor'' and the ``introduction of visual cues'' \cite{cruz2015wrongfully, cruz2016Semantic}. 

\subsubsection{Structural Metaphors}

Metaphors aid the understanding of a domain by grounding it in another one~\cite{lakoff1980Metaphors}.
While often associated with verbal language, metaphors can also be visual~\cite{pokojna2024language}. 
Cox, for example, argues that all data visualizations are metaphors that map data into forms to aid human understanding~\cite{cox2004artscience}. 
These metaphors can range from the ``parts to whole'' metaphor employed by pie charts to the use of lines to denote trends~\cite{zacks1999barslines}. 
Cox also proposes a metaphor $\leftrightarrow$ content spectrum, and argues that some metaphors have become conventional representations of data, like the node-link diagram to represent graphs~\cite{cox2004artscience}.  
Similarly, Ziemkiewicz and Kosara characterize visual metaphors broadly as the structures that form a framework to understand data~\cite{ziemkiewicz2008shaping}.
They found that participants reason about visual metaphors similarly to verbal ones, specifically to create mental models of the data~\cite{ziemkiewicz2009preconceptions}. 
Similarly, they found that the structure of the visualization metaphor can impact the implied dynamics~\cite{ziemkiewicz2010impliedDynamics} of points and their relations. In both cases, the form of the data can impact the ways that people reason about it. 
For example, they found participants perceived organizations presented in bubble charts as ``fun'' and 
``unserious'' while those depicted in waffle charts as ``bulky'' and ``regimented''. 
We propose that octopus maps similarly encourage conspiratorial interpretations of data, in line with other examples of conspiratorial visualizations that emphasize that ``everything is connected''~\cite{hannah2021conspiracy}.

In this paper, we focus specifically on how visual metaphors can impact the perception of the graph data that make up the octopus. The metaphor of the node-link diagram is dominant in graph visualization, having emerged from the 19th century with organizational charts, and in the 20th century with sociograms \cite{Correa2017HistorySocialNetwork}, as cited by Freeman~\cite{freeman}. These sociograms established many design conventions in modern node-link diagrams, such as the use of color and shape to denote attributes of nodes and edges~\cite{moreno1934whoshall}. The use of circular layouts likewise place important and connected individuals in the center of these visualizations. 

While node-link diagrams might be the standard metaphor for network data, networks have also been represented through visual encodings like rivers to represent dynamic hierarchies~\cite{bolte2021Split}, clusters of bubbles~\cite{Ukrop2010MetaphoricalVO}, or literal trees~\cite{kleinberg2001Botanical, deussen2004botanically}. Often, the goal of these alternate encodings is not sheer novelty's sake, but to re-express graph data in ways that make certain properties easier to detect. For example, treemaps more directly present hierarchical metaphors~\cite{ziemkiewicz2008shaping, johnson1991Treemap}, adjacency matrices can highlight clusters and cliques~\cite{mershack2019nodelink}, and edge bundling may emphasize or simplify trends in connectedness~\cite{wallinger2022EdgeBundling, vanderzwan2016Cubu, ozcan2011skeleton}. We note the visual similarity between the cartographically dispersed tentacles of octopus maps and \textit{flow maps}, which use the width of the edges connecting geographical locations to show flows across space---for example the migration of people~\cite{tobler1987experimentsmigrationmapping}.

We also draw attention to the role of metaphors in the space of science communication.
For example, Olson et al.~\cite{olson2019UserGuide} describe metaphors as ``key tools that are vital to science'' for visual communication in biology: ways of creating, reinforcing, and unifying conceptions of the field through shared terminology.
Olson et al. draw from philosophy and the study of linguistics to highlight characteristic properties of these metaphors, such as \textbf{expressiveness} (the degree to which a metaphor evokes the underlying phenomena), \textbf{paraphraseability} (the degree to which a metaphor can be replaced by equivalent or more specific terminology) and, perhaps most relevant to our work, an extension of Yablo's~\cite{yablo2000paradox} notion of \textbf{silliness} (the degree to which a metaphor contains aspects that are not true or relevant: for instance, the entity depicted as an octopus in the octopus map is almost never a literal sea creature that breaths underwater, has mile-long tentacles, or shoots ink). Part of the function of the octopus map is to map these ``silly'' characteristics onto a target (to portray it as a literal grasping entity with intentionality and purpose). Similarly, Pokojn\'{a} et al.~\cite{pokojna2024language} utilize Lakhoff and Johnson’s~\cite{lakoff1980Metaphors} framework of conceptual metaphors to characterize the types of visual metaphors used in scientific storytelling and data visualization. Conceptual metaphors are defined by mapping properties of one domain (e.g. the alien property of cephalopods) to another domain (e.g. some group of people or idea), and are classified based on how much or the kind of meaning they map. 
These span from solely graphical, called imagistic metaphors, to metaphors that map whole entities and structures from the source to the target, denominated structural metaphors.



\subsubsection{Pictorial Visualizations}

Pictorial elements in data visualization have a long and occasionally contentious record. Often, these visual components that do not directly encode data are disparaged under the (occasionally misleading) umbrella of ``chartjunk''~\cite{akbaba2021manifesto}. Yet, recent research has pointed to potential benefits for these kinds of ``embellishments''~\cite{cruz2016Semantic}. Especially for octopus maps, where the rhetorical goals include affective appeals and engagement, the inclusion of ``infographic'' styles of pictorial visuals~\cite{harrison2015infographic} has empirical benefits. For instance, improvements in memorability~\cite{bateman2010usefulJunk} and time spent lingering on the data~\cite{haroz2015isotype}. 

The pictoral elements most relevant to our work are the use of anthropomorphic and zoomorphic imagery in maps. There is a long history of using monsters or other animals in cartography, for instance to denote (or populate) unknown regions of the world~\cite{blake1999dragons,van2017hic} in  European maps from at least the Renaissance era onwards. In other cases, the entity itself is used as a metonym for a country as a whole, with well-known examples including the 16th century \textit{Europa Regina} (where the continent of Europe was depicted as a queen with various countries making up her constituent parts) and \textit{Leo Belgicus}~\cite{meurer2008europa} (where the Low Countries are depicted as a lion rampant, with geographic information depicted inside its ``body'').  Though many animals have been used to symbolize empires, countries, religions, and social and political entities, the octopus in particular is unique in its widespread and consistent use across eras, regions, and cultures~\cite{demhardt2023maps,Zanin2021Octopus,tyner1982persuasive}. The octopus has been used to incite fear, outrage, sympathy, disgust, and nationalism, all typically as part of a call to action.

For the communication of conspiratorial threat, the octopus itself is an especially apt symbol. The octopus and semiotically or psychologically related symbols like the medusa, kraken, or hydra~\cite{comandini1988octopus,schnier1956morphology,ronnberg2010book} have many negative cultural connotations that are useful for the designer of an octopus map. Cross-cultural fear of being captured or sucked in by the tentacles of an octopus recur in our dreams~\cite{comandini1991dream}. Each tentacle appears to move independently and with a mind of its own, a phenomena that Aristotle called \textit{poluplokon noema}: a ``multi-modal intelligence''~\cite{comandini1991dream} that is a non-human form of consciousness that both fascinates and repels~\cite{godfrey2016other}. The sinister connotations of the symbol of the octopus (or, an equivalent term seen in some of our maps, ``devilfish'') suggest a particular negative view of the data.

Octopus maps may be a special case where pictorial visualizations are especially useful for the (often nefarious) rhetorical goals of the designer. For one, the actual nature of the data (say, the specific graph structure) is often less important than an affective appeal (to, say, the omnipresence or seriousness of the threat). The disfluency or increased error rate observed in certain heavily pictorial or infographic-style visualizations may therefore be largely irrelevant, or even a ``beneficial difficulty''~\cite{hullman2011BenefittingInfoVis}---these maps may function more as a form of ``casual infovis''~\cite{pousman2007Casual} directed towards non-experts and reliant on confirming existing attitudes towards the data they contain.

\section{The Octopus Map and its Subspecies}
\label{sec:examples}


In this section, we explore a set of historical examples of octopus imagery in maps to both better define our genre of interest as well as to portray the scope, longevity, and diversity of this genre. As prior work illustrates~\cite{correll2024body}, modern visualization practices are often highly influenced by historical precedents, and analyses of historical examples can provide inspiration or guidance for rhetorical goals or design problems that recur across centuries. Our examples are sourced variously from the last author's personal corpus (amassed from 2022 onwards) of 81 examples encountered in the wild, as well as from museum collections~\cite{pjmode,barber2010magnificent}, blogs~\cite{vulgararmy}, and magazine articles~\cite{meier2017octopus,nagy2015bizarre,ottens2017octopus,young2016octopus}. We were skeptical of the existence of satisfactorily complete or thematically exhaustive corpora, and so selected our corpus of examples for breadth and variety rather than representative completeness. Octopus maps are, in a somewhat contradictory fashion, ubiquitously produced but rarely preserved: per Barber et al.~\cite{barber2010magnificent}, ``the survival of [propaganda maps] is extremely rare as despite being mass produced for a large audience, they were rapidly discarded and destroyed''. Likewise, there are many examples of maps that do not depict literal octopuses but are relevant in more borderline or implicit ways (see \S\ref{sec:edges}). We note also that, while our corpus contains both historical and contemporary examples of octopuses representing entities as varied as the ``Evil Literature Menace'', Los Angeles, and Scientology, our decision to reproduce in this paper only images in the public domain or open archives bias the figures in this paper towards examples from the late 19th and mid 20th century: the earliest example reproduced in this paper dates from 1877 (\autoref{fig:russia}).

\begin{figure*}
    \centering
    \begin{subfigure}[t]{0.415\textwidth}
        \centering
        \includegraphics[width=\textwidth]{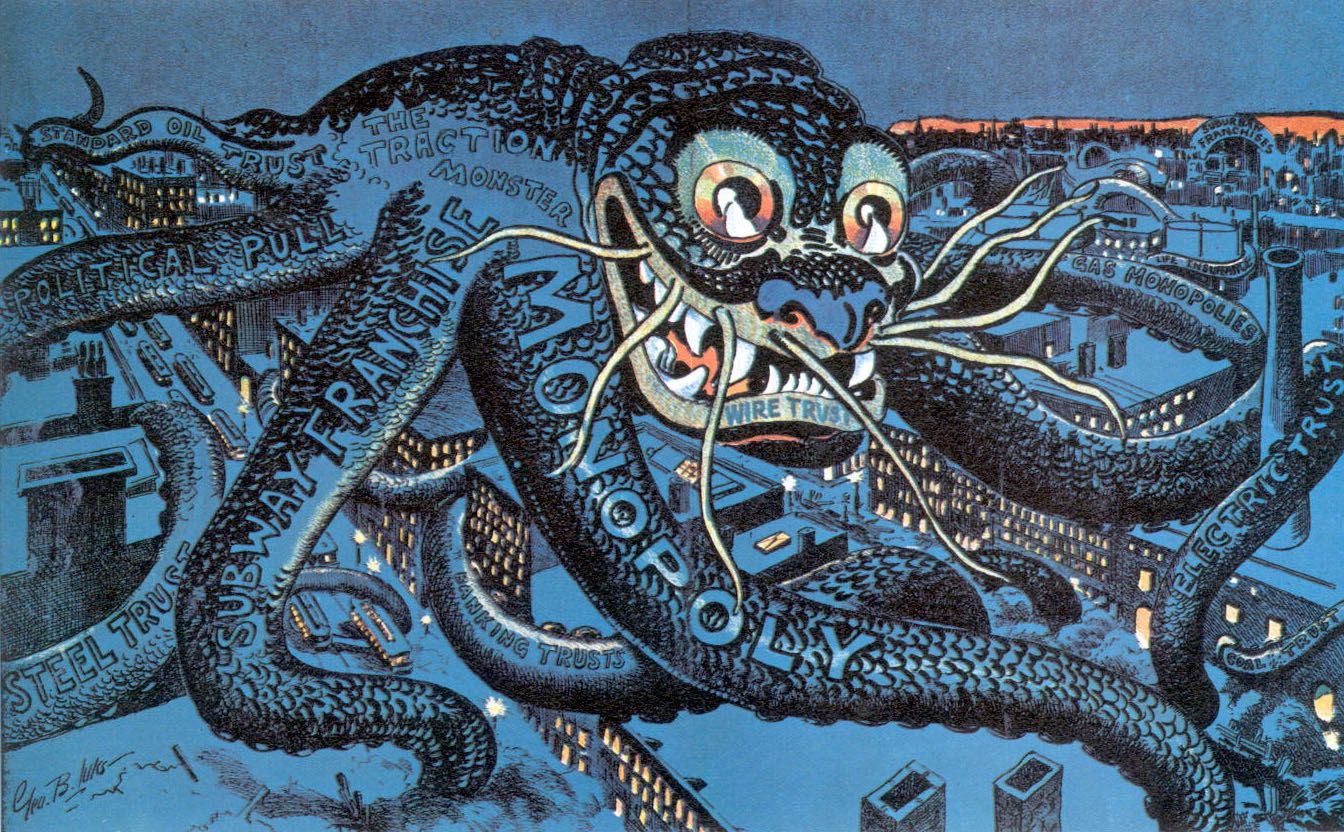}
        \Description{A political cartoon from 1899 titled "The Menace of the Hour."  The menace illustrated is an octopus labeled the "Traction Monster" which represents the new electric subway system in New York City. The octopus has tendrils labeled standard oil trust, political pull, steel trust, subway franchise, banking trusts, electric trust, and gas monopolies. The octopus's lower lip is labeled wire trust. The implication is that the New York City electric subway system franchise controls or influences the industries and trusts listed on the octopus tentacles.}
        \caption{From 1899, ``The Menace of the Hour'' depicts the new electric subway system in New York City as a monopolistic octopus}
        \label{fig:subway}
    \end{subfigure}
    ~
    \begin{subfigure}[t]{0.425\textwidth}
        \centering
        \includegraphics[width=\textwidth]{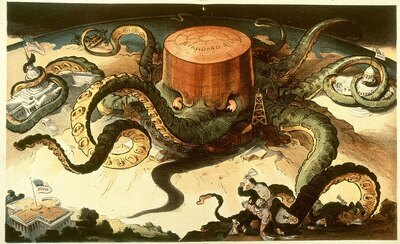}
        \Description{A political cartoon from 1970 depicting an octopus labeled "Oil industry" with its tentacles grabbing hold of the White House, Treasury Dept., U.S. Capitol, foreign policy and the gas, coal, and nuclear energy industries.}
        \caption{A 1904 illustration by Udo Keppler depicting the Standard Oil company as an octopus grasping for other industries and centers of government, including the U.S. Capitol and the White House}
        \label{fig:oil}
    \end{subfigure}
    \caption{Political cartoons using the octopus motif, in both cases implying monopolizing entities as octopuses acquisitively grasping for control.}
    \label{fig:cartoons}
\end{figure*}

Not all usages of octopus iconography are octopus \textit{maps}. For instance, political cartoons often employ an octopus to represent a variety of concepts, and the resulting cartoons may lack specific cartographic meaning, beyond the labelling of tentacles to indicate connected ills (\autoref{fig:cartoons}). While we collected examples of these non-cartographic octopus maps in order to better assess the design techniques that can emphasize octopus-like rhetoric, in this paper we focus mostly on the use of the octopus motif in persuasive cartography.

We organize the examples in this section by first considering what we think of as canonical examples (where a country in a geographic region is portrayed as an octopus, with tentacles extending towards or around its neighbors). We then discuss cases that expand on this definition (for instance when, rather than a country, an ideology is the center of the octopodal body) or problematize it (as when octopus-like rhetoric is implicitly rather than explicitly conveyed).

\subsection{The Canonical Octopus Map}
\label{sec:canon}

\begin{figure}
    \centering
    \includegraphics[width=0.5\textwidth]{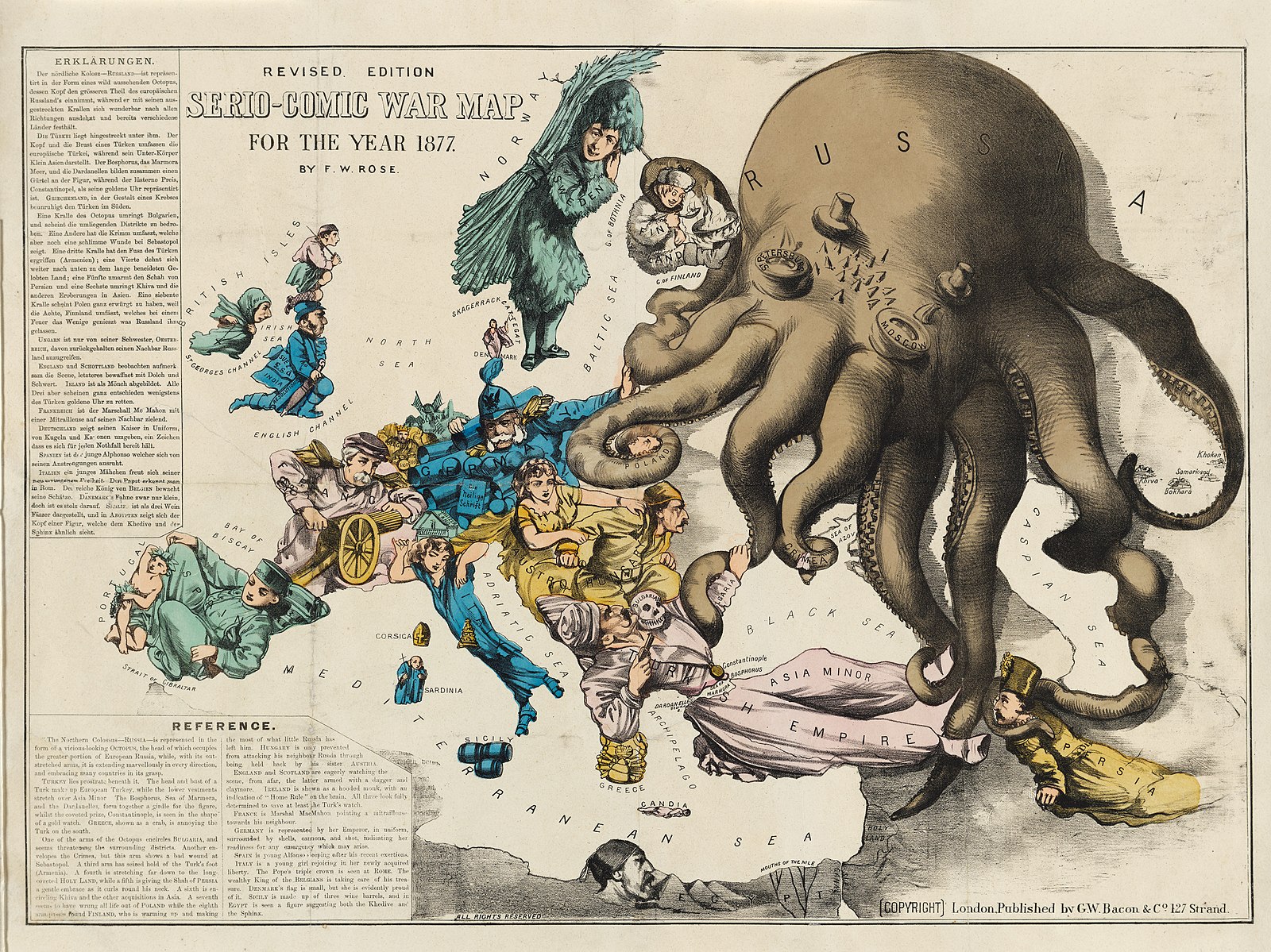}
    \Description{An illustrated map from 1877 depicting a portion of the world map limited to Europe with small portions of North Africa and Southwest Asia. Included countries are drawn as people except for Russia, which is depicted as an octopus battling the Turkish Empire and grasping for various European and Southwest Asian lands, including Finland, Poland, and Persia.}
    \caption{\textit{The Serio-Comic War Map for the Year 1877} by Frederick W. Rose, published shortly after Russia attacked the Ottoman Empire in response to the Turkish massacre of Christian Bulgarians. Here Russia is depicted as the expansionist octopus, battling the Turkish Empire and grasping for various other countries, including Finland, Poland, and Persia.}
    \label{fig:russia}
\end{figure}

As mentioned, we consider the canonical octopus map to be a map where a country (or a synecdoche or other metonym for that country, such as its leader or national personification) in a geographic region is depicted as an octopus, with the tentacles of this octopus extending to neighboring countries or regions. These tentacles usually denote some negative relation such as actual or intended conquest, control, or other subversion of victims. For instance, one of the earliest known examples of an octopus map was published by Fred W. Rose in 1877 during the Russo-Ottoman War (\autoref{fig:russia}). The map depicts Europe, with most countries represented by human figures, while Russia is depicted as an octopus with tentacles outstretched, reaching for its neighbors. 

War, as a quintessential and dramatic adversarial relation between countries, is a rich source of octopus maps. In addition to the Russo-Ottoman example above, octopus maps are present for many major 19th and 20th century conflicts, including the Russo-Japanese War, World War I, World War II, the Vietnam War, and the Cold War era~\cite{mode2017maps}. In many examples, victimized countries are not anthropomorphized as in Rose’s map, but are instead left as cartographic areas or represented by flags. For instance, the 1882 American political cartoon ``The Devilfish in Egyptian Waters'' shows John Bull (the national personification of the United Kingdom) with octopus-like tentacles (each with a human hand) grasping for colonial acquisitions around the world~\cite{akengin2022assessment} that are more iconic than geographic. The octopus is the central figure in these examples, emphasized through color, size and symbology.

\begin{figure*}
    \centering
    \begin{subfigure}{0.475\textwidth}
        \centering
        \includegraphics[width=\textwidth]{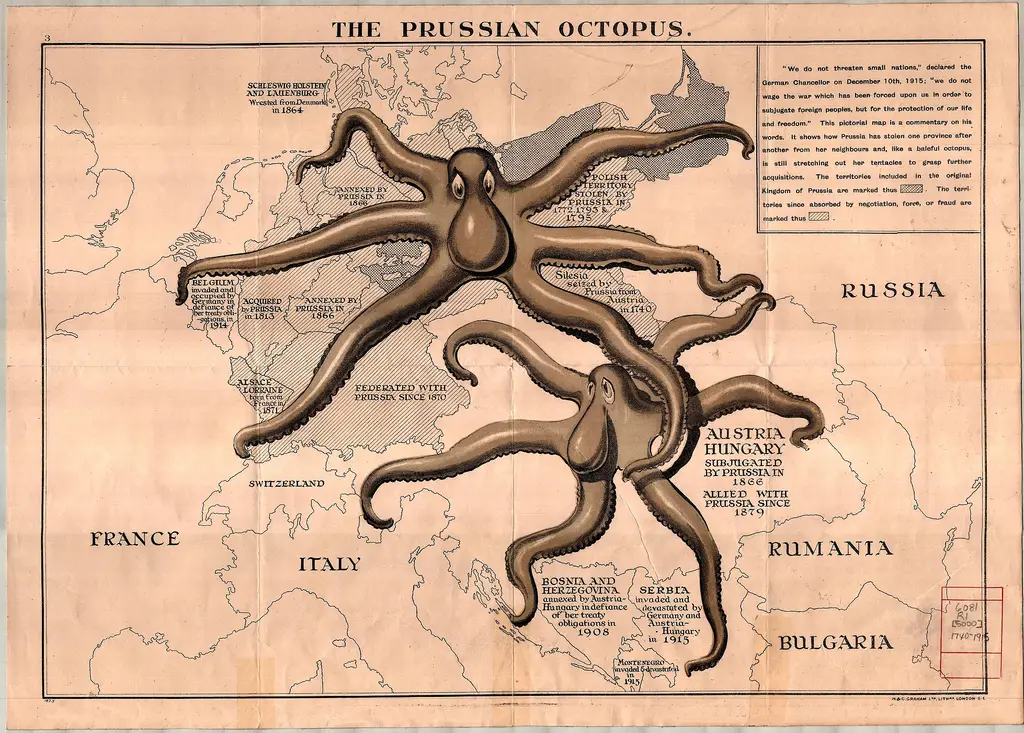}
        \Description{A partial map of Europe from 1916 depicting Prussia as a large, central octopus with a tendril over Austria-Hungary, also represented as an octopus. The tentacles of both octopi reach out to various countries in the region. The map notes that on December 10th, 1915, the German Chancellor declared, ‘we do not wage war which has been forced upon us in order to subjugate foreign peoples, but for the protection of our life and freedom.’ The Prussian and Austria-Hungary octopi are depicted with their listed territorial conquests in the map, contradicting the Chancellor's words.}
        \caption{A circa 1916 octopus map from an unknown author titled ``The Prussian Octopus'' depicting Prussia as a central octopus subjugating neighboring regions. The date and context of the acquisition of each bit of territory is labelled \textit{in situ}. Austria-Hungary is depicted as a (perhaps reluctant) subsidiary octopus with its own acquisitions, but still under the watchful eye (and tentacle) of Prussia.}
        \label{fig:prussia}
    \end{subfigure}
    ~
    \begin{subfigure}{0.475\textwidth}
        \centering
        \includegraphics[width=\textwidth]{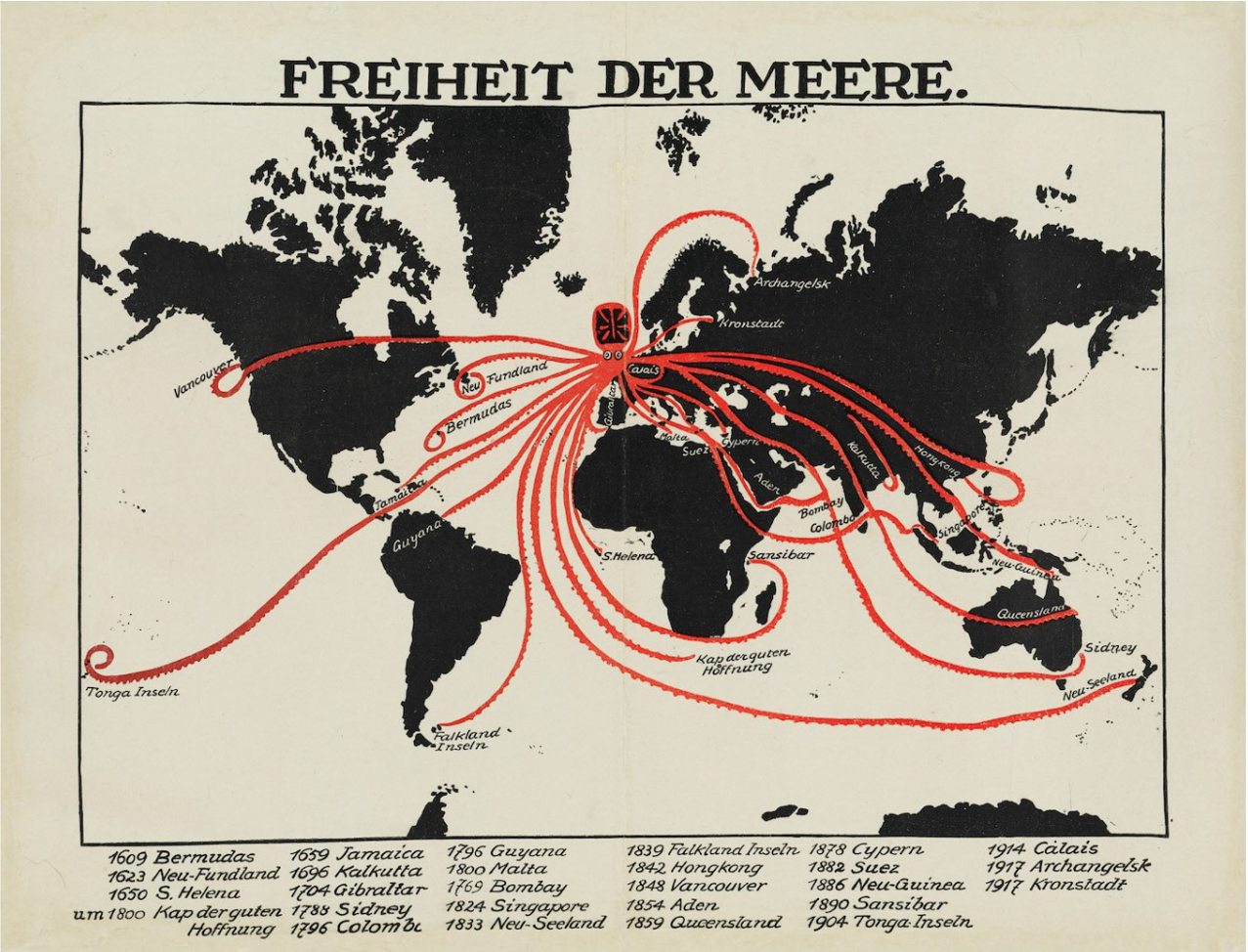}
        \Description{Depicts a full world map from 1917 titled 'Freedom of the Seas,' in German  with England as a red octopus with tentacles reaching out to countries around the world. Countries are listed in German below the map with dates associated with control or colonization by England.}
        \caption{A circa 1917 octopus map titled ``Freedom of the Seas''~\cite{freiheit} from an unknown author depicting England as a central octopus with tentacles extending to its far-flung island colonies. Other versions of this poster occasionally include the subtitle ``England the bloodsucker of the world''~\cite{vulgararmy}. The caption includes the date of control of each island.}
        \label{fig:england}
    \end{subfigure}
    \caption{Two octopus maps from opposite sides of the first World War, each using similar designs and data to cast their opponent as an octopus. On the left, Prussia's century of territorial change casts it as an acquisitive octopus, with the Austro-Hungarian empire a subordinate but nonetheless acquisitive partner in crime. On the right, English colonial acquisitions across the world from the 7th century onward are cast as part of a plan of control of the world's oceans.}
    \label{fig:ww1}
\end{figure*}

\begin{figure}[t]
    \centering
    \includegraphics[width=0.35\textwidth]{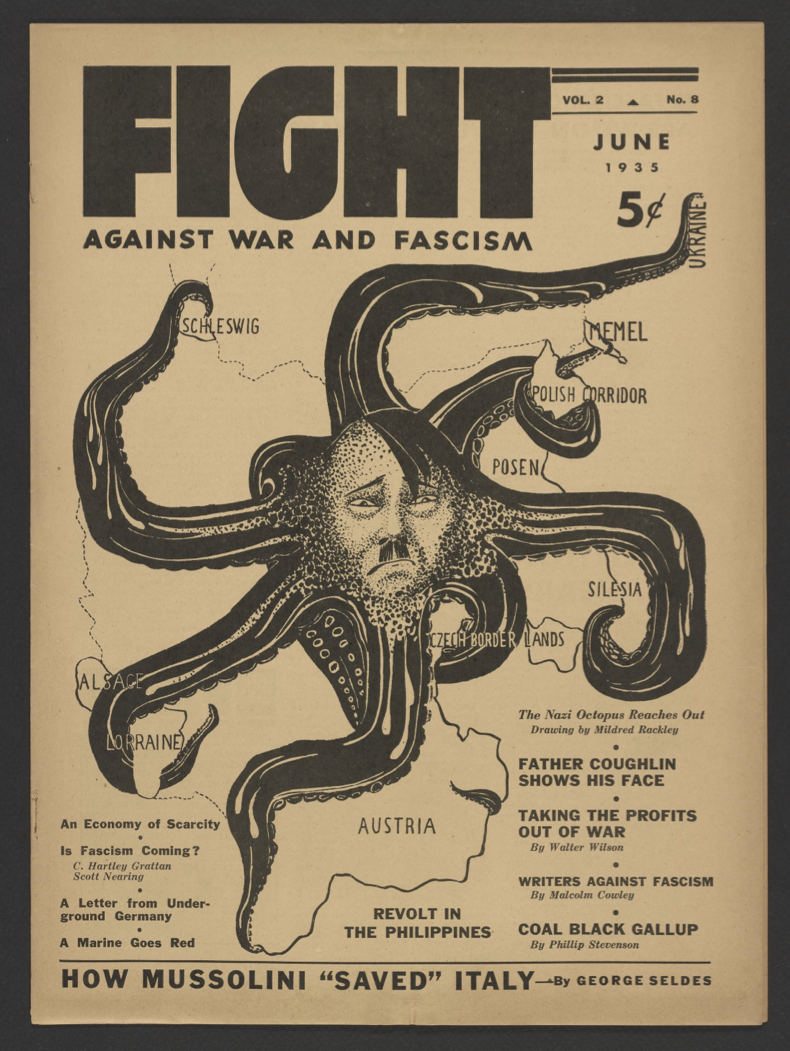}
    \Description{A large octopus whose head is drawn as Adolf Hitler with tentacles reaching out to various locations in European. The title is 'fight' in all capital letters. This is the cover of a 1935 antifascist magazine.}
    \caption{The cover of a 1935 antifascist magazine depicting Adolf Hitler as a fascist octopus with tendrils around neighboring lands~\cite{fascism}. Note that, at the point of this publication, many of the acquisitions noted here had yet to occur (for instead, Austria in the \textit{Anschluss} of 1938, or Alsace-Lorraine as part of terms of the armistice after the fall of France in 1940).}
    \label{fig:hitler}
\end{figure}

We note that octopus maps can be produced by both sides of a conflict, using similar visual rhetoric to produce opposing arguments. For instance, during World War I, an Entente propaganda map shows a Prussian octopus with an Austro-Hungarian junior partner (\autoref{fig:prussia}), with Prussian territorial gains from 1772 onwards used as evidence of Prussia's goal of expansion and dominion. But a contemporaneous map from the Central Powers shows the British Empire's acquisition of overseas territory from 1609 onwards as part of a campaign of control over the oceans (\autoref{fig:england}). This battle of maps also occurred in the Second World War---while depicting Germany or the ideology of fascism as an octopus was common in Allied propaganda (as in \autoref{fig:hitler}), there are also examples from Axis powers (or countries occupied by Axis powers) severing the tentacles of the American ``Dollar Octopus''~\cite{dollarpoliep} or suggesting that Axis actions were part of a ``methodical'' plan to ``amputate'' the tentacles of the English imperial octopus~\cite{amputation}.

\subsection{The Ideological Octopus}
\label{sec:ideology}

\begin{figure}[t]
    \centering
    \includegraphics[width=0.35\textwidth]{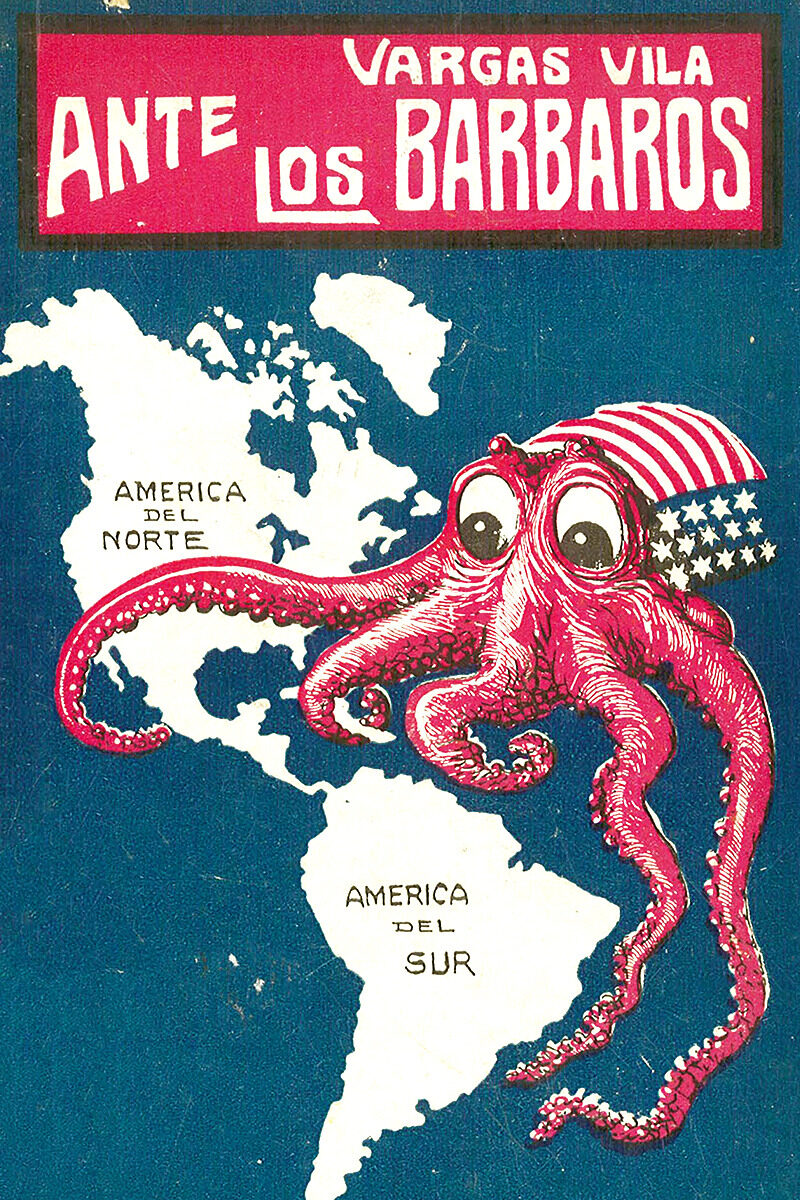}
    \Description{Shows the western hemisphere with a large octopus looming over it. The octopus has an American flag on its head and is reaching out to Central and South America.}
    \caption{The 1930 cover of \textit{Ante los Bárbaros} (``Before the Barbarians'') by J. M. Vargas Vila, an anti-imperialist work accusing the United States of using the First World War as cover to continue its imperialist ambitions in the western hemisphere without European interference.}
    \label{fig:usa}
\end{figure}

\begin{figure}
    \centering
    \includegraphics[width=0.5\textwidth]{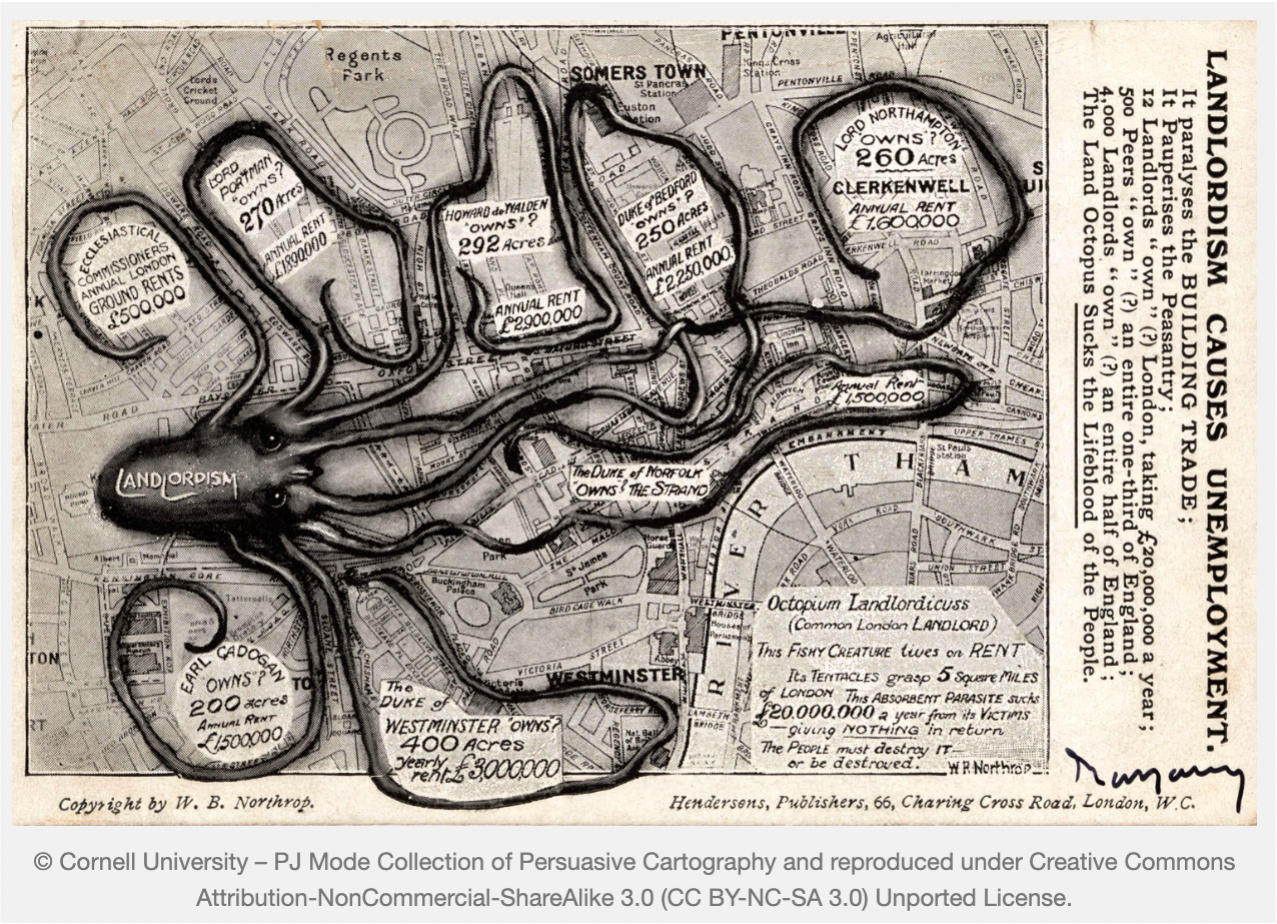}
    \Description{A political cartoon of a map of London with an octopus laid over so that its tendrils encircle different neighborhoods. Each of these neighborhoods lists a landlord with the number of acres they own and the annual rent they make from their properties.}
    \caption{1909 political cartoon by W. B. Northrop titled ``Landlordism Causes Unemployment,'' showing the octopus of landlordism strangling London, with particular neighborhoods annotated with the specific area and rent amounts controlled by individual landowners. Image courtesy Persuasive Maps: PJ Mode Collection, Cornell University Library.
}
    \label{fig:landlordism}
\end{figure}

The tentacles of the octopus map do not exclusively denote martial conquest.
Other maps cast colonialism, imperialism, and other forms of domination or control as tentacles.
Social movements have also leveraged the looming figure of the octopus to  decry monopolies, industries, specific companies, and trade practices (\autoref{fig:cartoons}).
In Central America, the United Fruit Company was commonly referred to as \textit{el pulpo} (``the octopus'').
United Fruit established a vast network of infrastructure supporting the banana market from the 1870s to the mid-20th century.
While no longer in operation, the company remains a symbol of American imperialism and greed owing to its long history of corruption, violence, and environmental destruction in Central and South America ~\cite{bucheli2005bananas, bucheli2008multi}.
The U.S. imperialist octopus therefore appears in octopus maps of Central American anti-imperialists (\autoref{fig:usa}).

W.B. Northrop presents another example of this ideological octopus by portraying landlordism as the cephalopod's body with its tendrils physically surrounding the areas of the city owned by notable English aristocrats (\autoref{fig:landlordism}). 
Specifically, Northrop used this visual motif to criticize British inequality, inspired by land reform proposals of  liberal politician David Lloyd George. 
The octopus portrays landlords as the \textit{Octopium Landlordicus}, or the ``Fishy (Rent Eating) Creature''.

In a more recent example from the International Journal of Cartography, Zanin and Lambert~\cite{Zanin2021Octopus} examine design choices for an octopus map of arms sales by the NATO military alliance. They assess the impact of factors like map projection, color choice, and circle proportions, while also clearly stating the rhetorical purpose of the map to call out the scale and impact of arms sales from NATO countries and question existing hegemonic structures.

\begin{figure}
    \centering
    \includegraphics[width=0.475\textwidth]{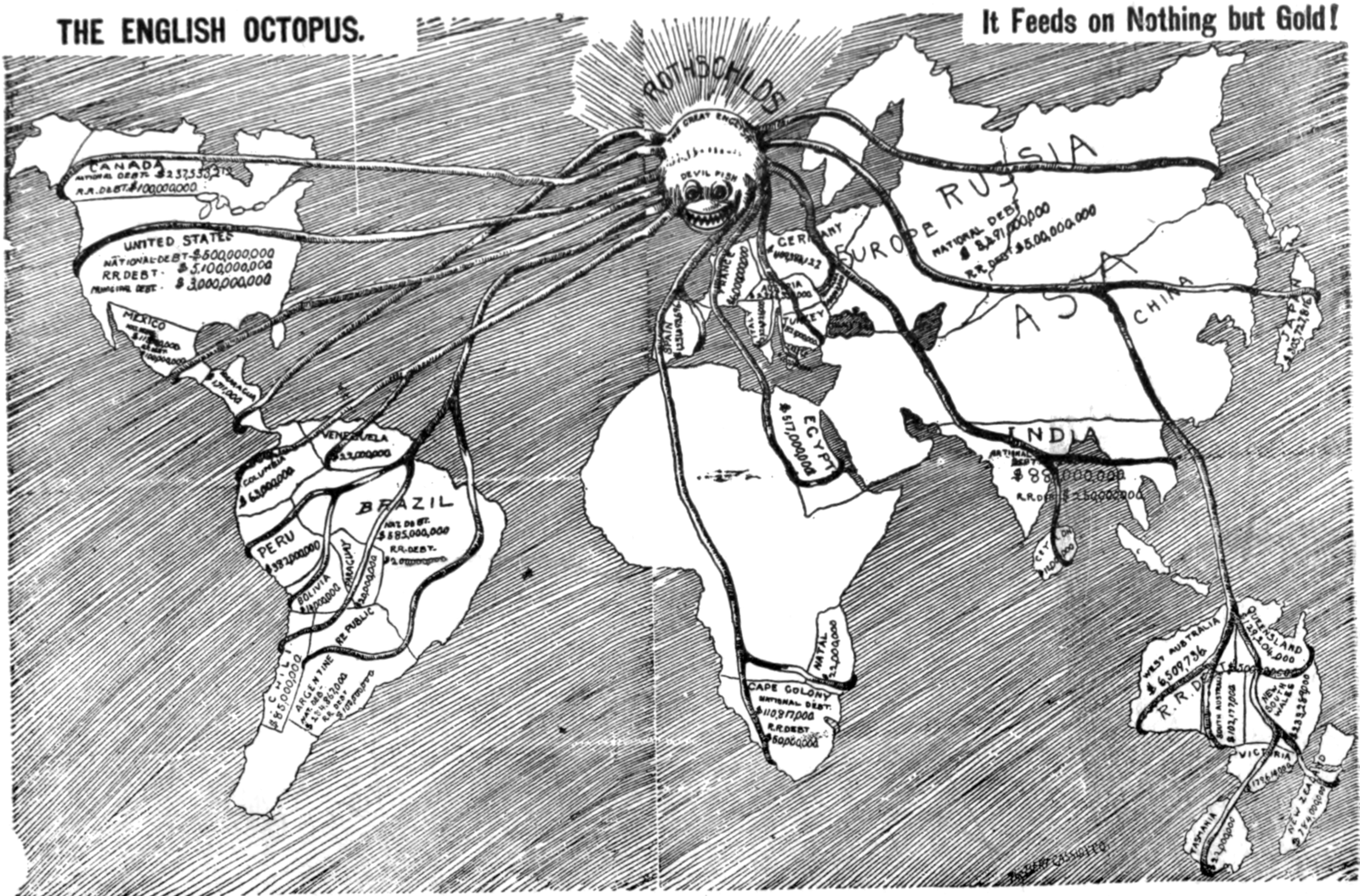}
    \Description{A drawing of a world map from 1894 with an octopus labelled 'The Rothschilds' with tentacles reaching out to many regions across the world. The title is 'The English Octopus' and a secondary title on the top right says, 'It Feeds on Nothing but Gold!'
    The Rothschilds were a prominent Jewish family highly associated with the 18th and 19th century European banking.}
    \caption{1894 depiction of the Rothschilds (a prominent Jewish family highly associated with the 18th and 19th century European banking) as an octopus from a pamphlet written by William Hope Harvey in support of moving away from the gold standard and towards ``bimetalism''.}
    \label{fig:rothschild}
\end{figure}

However, these ideological octopus maps can also dehumanize religious or ethnic groups. For instance, early editions of the influential antisemitic text \textit{The Protocols of the Elders of Zion} had covers depicting a Jewish octopus, snake, or spider encircling the globe~\cite{bronner2003rumor}, a motif that recurs in Nazi propaganda decades later. Fear of Jewish financial control is also part of the implicit rhetorical work in \autoref{fig:rothschild}: while the title is ``The English Octopus'', a smirking octopus representing the Rothschild family obscures the already simplified and distorted cartography of the map. Given the history and ubiquity of octopus motifs in antisemitic work, there are debates on the extent to which the octopus, \textit{per se}, represents an intrinsically antisemitic symbol~\cite{thunbergocto}.



\subsection{The Implicit Octopus}
\label{sec:edges}

The octopus has no monopoly on persuasive cartography. Many of the rhetorical goals of an octopus map (such as showing centrality or connectedness, or conveying menace) can be accomplished through other means. We were especially interested in how techniques or designs common to octopus maps could appear in other forms of maps or visualizations, and so implicitly or subtly convey what octopus maps might show overtly.

Some early examples of octopus map-like structures are route maps like the 1837 railway traffic map by Sir Henry Harness for the Irish Railway Commissioners, in which passenger flow is shown as a network overlaid on a map (\cite{harness1837atlas}, as cited by Robinson~\cite{robinson1955henrymaps}). In this map, thick lines emanate from Dublin to convey the city's centrality and the ability to reach even the most remote parts of Ireland, although the map does not explicitly sensationalize or propagandize these flows.
We note that many of the visual features we observe in octopus maps can be beneficial for reading route maps and other flow diagrams. For example, curved lines have been shown to be more effective than straight lines in flow maps, and are often preferred by users~\cite{bar2006humans,silvia2009do}. Central nodes included in flow maps, rather than areas, can also lead to lower error rates in viewer interpretation~\cite{jenny2016design}.

Other implicit examples rely heavily on network diagrams, but lack cartographic elements. One example is \textit{They Rule}, an interactive visualization showing the interlinked boards of the top 100 US companies in 2021 ~\cite{on2002they}. The resulting graphs are often interconnected and expansive, creating an octopus-like interpretation of the many links between major corporations. Other edge cases include visualizations that depict an oppressive or controlling force without incorporating the same central, directional force of an octopus map. Although the authors of this work were divided on this particular example, there are also implicitly octopodal elements in the ``Pyramid of the Capitalist System''~\cite{pyramid}, a 1911 poster made for \textit{The Industrial Magazine} and since then remixed, repurposed, or otherwise reused for anti-capitalist rhetoric into the modern age. In the poster, a bag of money labelled ``Capitalism'' is at the top of pyramidal structure of social classes: from the government leaders who ``rule you'' to the military who ``shoot at you'' and then, finally, the workers and farmers crushed under the weight at the bottom. The resulting graph structure is implicit---various layers are associated in a hierarchy with a central figure on top. However, the rhetorical message of a central controlling force is similar to more explicitly octopodal examples.

It is these implicit ``octopus-like'' designs, and the sometimes fuzzy boundary between overtly sensationalized octopus maps and more allegedly neutral (but still potentially octopodal) designs, that motivated us to decompose and then evaluate the visual and structural elements of the canonical octopus map in the following sections.

\subsubsection{The Benevolent Octopus}
\label{sec:benevolent}

\begin{figure*}
    \centering
    \begin{subfigure}[h!]{0.45\textwidth}
        \centering
        \includegraphics[width=\textwidth]{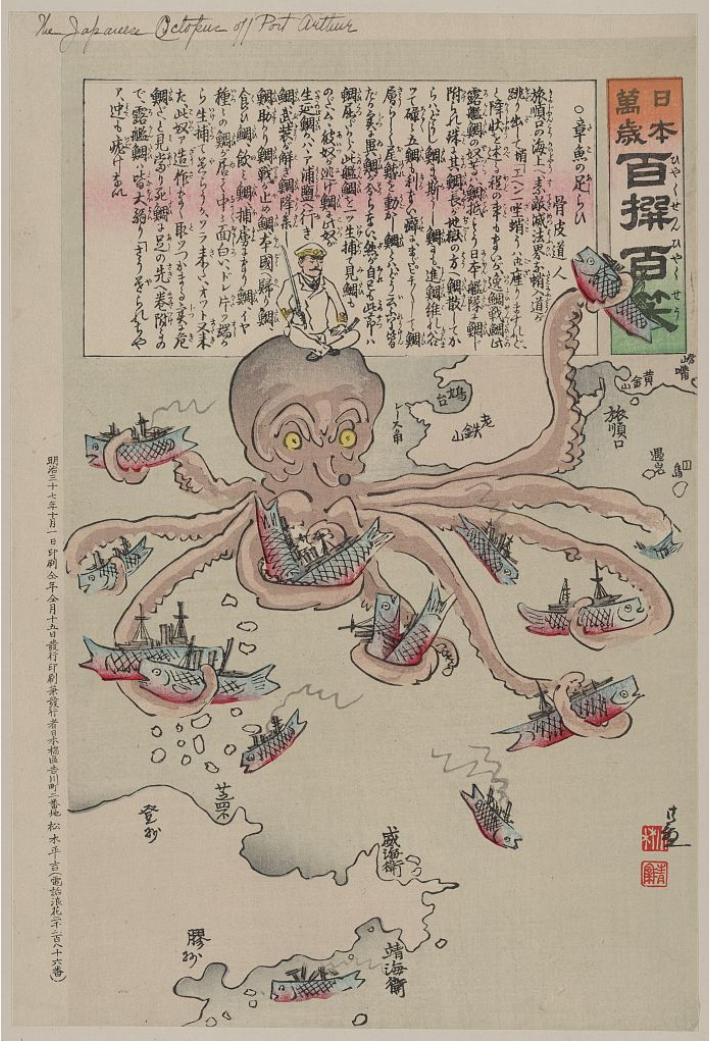}
        \Description{A print from 1904 depicting an octopus off the coast of Japan. The print celebrates a Japanese victory in the Russo-Japanese war. The octopus grasps several Russian ships which are drawn to resemble fish.}
        \caption{}
        \label{fig:japan}
    \end{subfigure}
    ~
    \begin{subfigure}[h!]{0.45\textwidth}
        \centering
        \includegraphics[width=\textwidth]{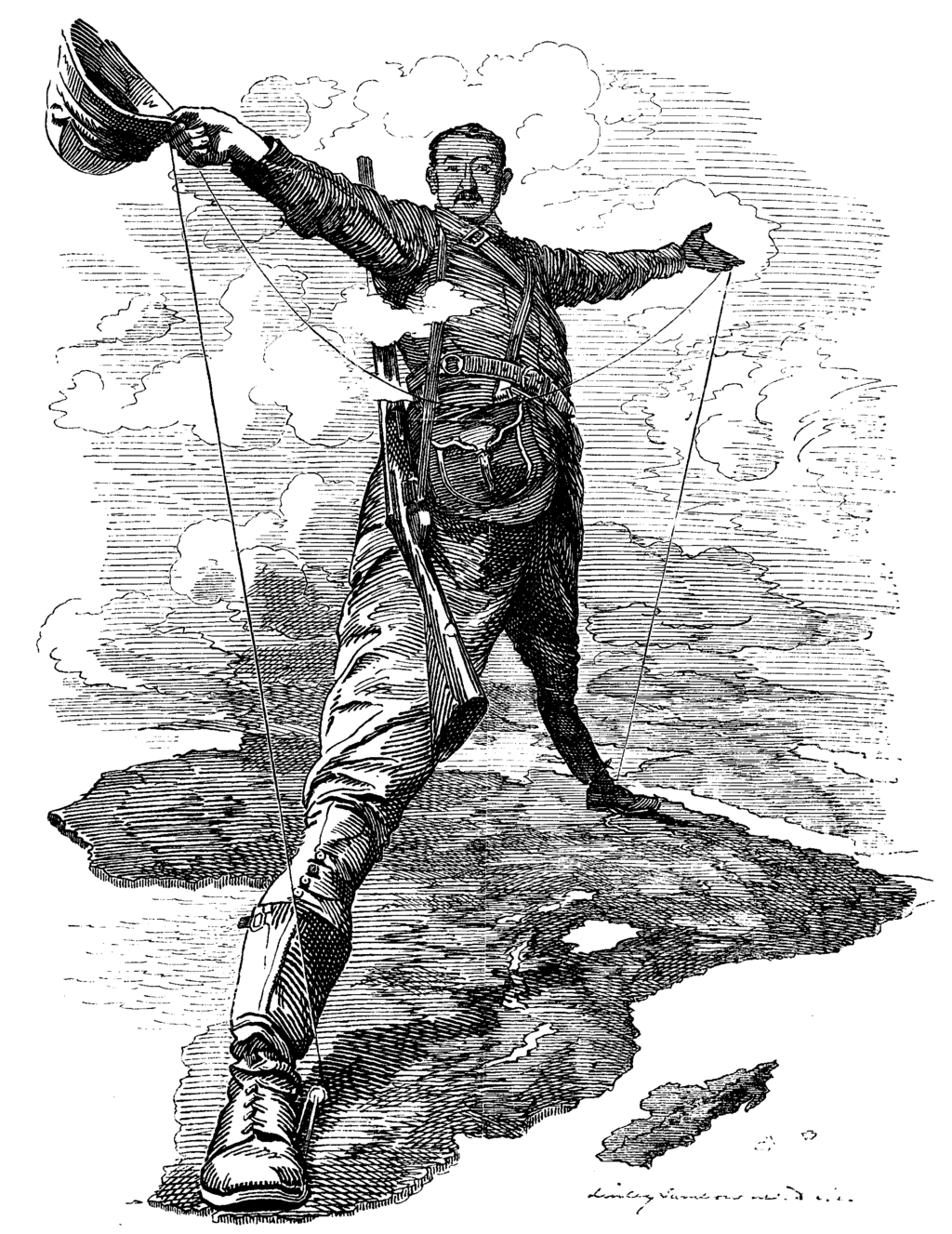}
        \Description{A political cartoon showing a large male figure holding a line across the African continent. This represents the proposed rail and telegraph line connecting British colonies in Africa.}
        \caption{}
        \label{fig:rhodes}
    \end{subfigure}
    \caption{(a) 1904 print by Kiyochika Kobayashi of a Japanese victory from the Russo-Japanese war, with a victorious admiral sitting atop an octopus holding fish-like captured Russian ships in its tentacles. (b) 1892 political cartoon published in Punch magazine and illustrated by English cartoonist Edward Linley Sambourne depicts British business magnate Cecil Rhodes as a giant standing over the continent holding a telegraphic line, referencing his "Cape to Cairo" rail and telegraph line efforts to connect most of the British colonies in Africa.}
    \label{fig:benevolent}
\end{figure*}

While many examples of the octopus map portray the octopus as a villain, some depict it as a benevolent force (\autoref{fig:benevolent}). These examples are less common, but they represent an alternative version of the form worth discussing. In these cases, the long reach and solid grasp of tentacles can be seen as protective or even triumphant. For instance, a Japanese map from the Russo-Japanese war (\autoref{fig:japan}) casts an octopus as a symbol of Japanese victory and a tool for the expulsion of the enemy.

However, given the generally negative associations of the octopus, other maps that seek to portray omnipresent connection often employ the implicit structures we mention in the previous section, producing maps that have much in common visually or stylistically with octopus maps while avoiding the negative connotations of the octopus motif. For instance, in Edward Linley Sambourne's 1892 caricature of British business magnate Cecil Rhodes (\autoref{fig:rhodes}), the prospect of Rhodes' proposed but never completed ``Cape to Cairo'' rail and telegraph line is represented by Rhodes himself standing across the African continent holding his line, connecting British colonies in Africa.

As a second example, B. Milleret's~\cite{reverseFrance} 1931 illustration is a particularly striking example of using visual metaphors associated with octopus maps to depict (purportedly) benevolent rather than threatening relationships. Commissioned by the French Army, the caption reads ``It is with 769,000 people that France ensures peace and the benefits of its civilization to 60 million native peoples.'' While other maps in this paper have used far-flung colonial possessions to build visual arguments for a country as a threatening or oppressive force, in this map France is a depicted as a central ``sun'', with tentacular rays of light reaching out to each of its colonial possessions throughout the world. 

While less overtly ideologically laden, other forms that convey octopus-like arguments or designs without intending threat or proposing action are the route maps put out by airlines. Central hub airports are connected by arcs to destinations across the globe representing available flights. The resulting map gives the impression of a benevolent omnipresence: that the airline goes wherever you need to be.

\section{Dissecting the Octopus Map}
\label{sec:parts}
While our historical examples have a variety of different styles, data, and rhetorical targets, we maintain that the overall \textit{visual arguments} embodied by these maps have commonalities. Not all juxtapositions of graphs and cartography are octopodal in nature: creating sensationalized octopus-like impressions (of, say, a nefarious controlling body with multiple independent arms of attack), seems to involve a confluence of implicit or explicit choices in graphical structure and representation.

After several rounds of debate and iteration among the co-authors, we propose the following central components of a prototypical octopus map. We also describe some of the relevant visual and rhetorical characteristics, specifically in relation to how the octopus mimics elements of conspiratorial thinking. We choose these properties because they allow us to analyze octopus maps via two complementary lenses: through the \textit{structure of the data} underlying the map (in most cases, as a network of connections occurring in a spatial area) as well as the \textit{visual techniques layered atop this data} (such as the decision to depict edges as literal tentacles). The confluence of these components, we hold, lends itself to the particularly conspiratorial nature of these sensationalized maps. However, we also maintain that some of these elements can act to produce ``octopus-like'' interpretations of maps that otherwise stray from the canonical form we lay out in \autoref{sec:canon}. Our user study (\S\ref{sec:experiment}) specifically explores the extent to which these various individual components can be elided or altered in maps while still contributing to overall octopodal arguments. 

\newcommand{\cardsize}{0.15\columnwidth}
\newcommand{\cardparasize}{0.85\columnwidth}

\vspace{1em}
\noindent
\begin{minipage}{\cardsize}
\includegraphics[width=\textwidth]{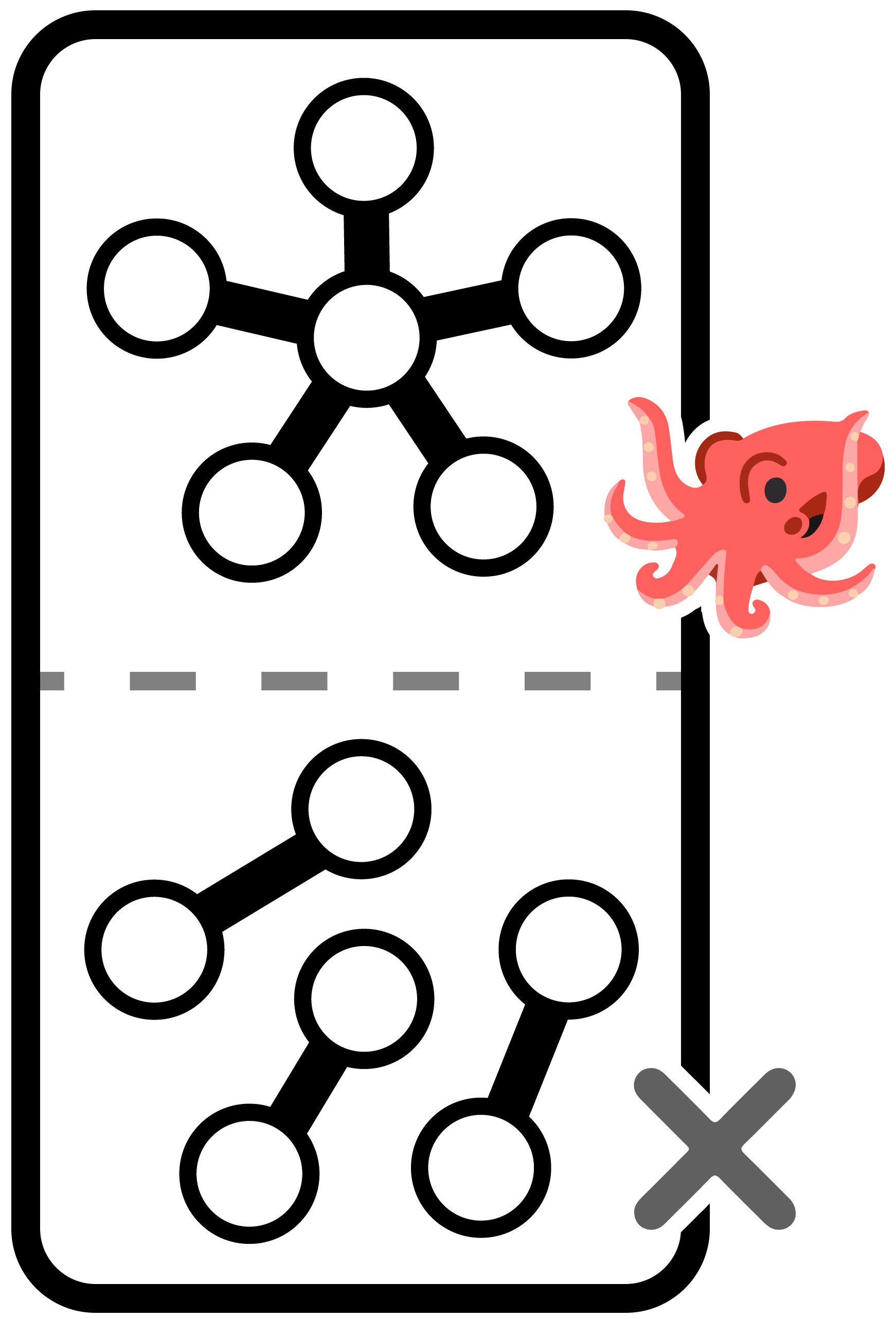}
\Description{A card with two network diagrams, the top one representing our concept of centrality by showing all nodes connected to the center, with the bottom one showing disconnected pairs.}
\end{minipage}
\begin{minipage}{\cardparasize}
    \textbf{Centrality}: \textit{while there might appear to be a number of disconnected or diffuse threats, they are in fact connected to a single \textit{central body}}. Assigning blame for diverse ills to a single cause is the \textit{sine qua non} of conspiratorial thinking~\cite{douglas2023whatareconspiracy,douglas2016someone,douglas2018whyconspiracy,imhoff2014speaking}. While centrality is straightforward in octopus maps---there is a literal octopus body with tentacles---it is also a property of graphs in general, and visual elements like alignment, layout, and visual motifs can also amplify perceived centrality~\cite{muehlenhaus2013design}.
\end{minipage}


\vspace{1em}
\noindent
\begin{minipage}{\cardsize}
\includegraphics[width=\textwidth]{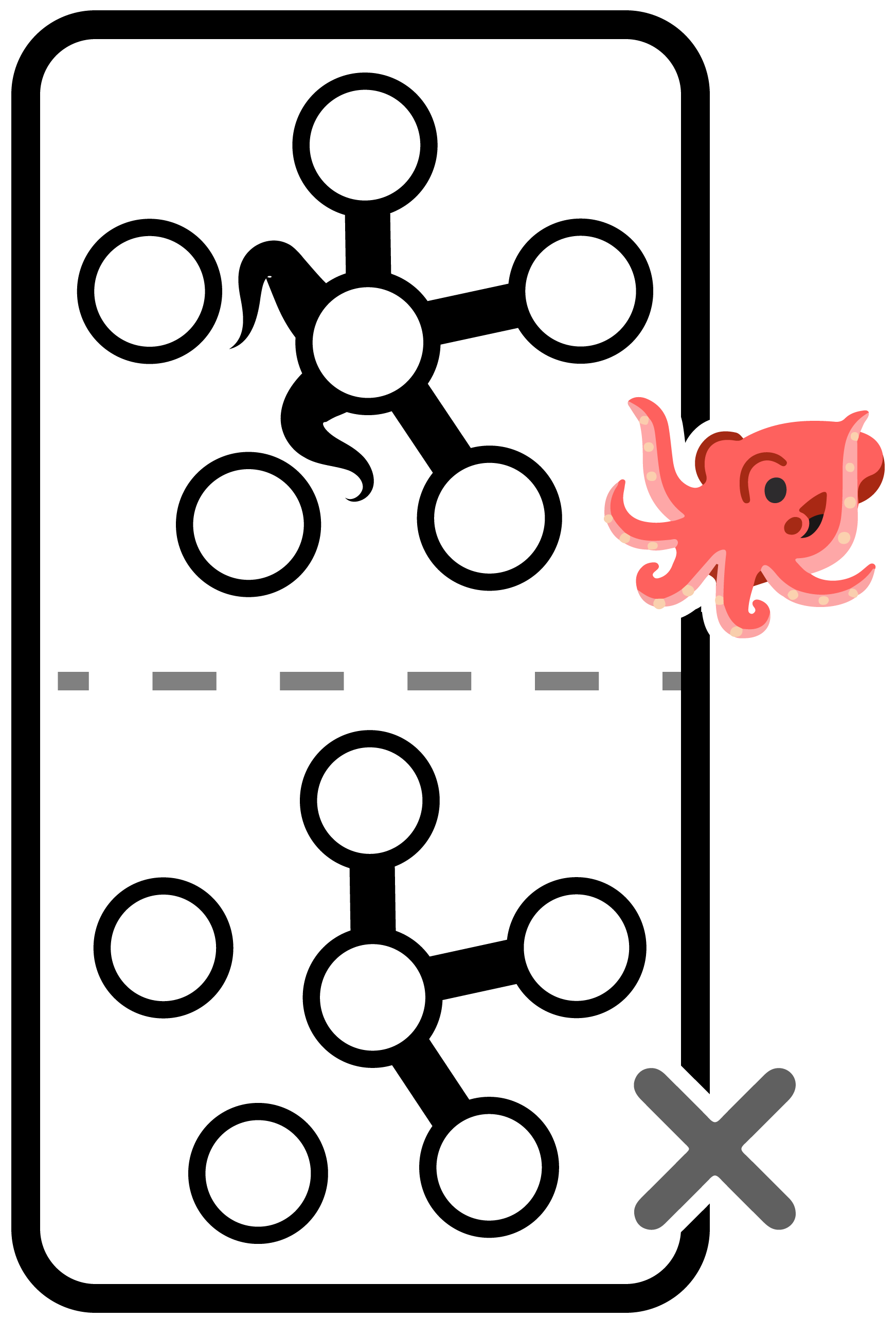}
\Description{A card with two identical network diagrams, except that the top one represents our concept of tentacularity by showing nodes with wavy prospective edges reaching out towards disconnected nodes.}
\end{minipage}
\begin{minipage}{\cardparasize}
  \noindent\textbf{Tentacularity}: \textit{this central body possesses a number of ``tentacles'': directional threats that may appear to have some degree of independent control while still following the will of the central body}. For octopus maps, this agency can be conveyed through visual dynamism in the tentacles, with the idea that each independent limb represents a threat and is seeking to control even if the other limbs are engaged elsewhere. Implicitly, tentacularity can be conveyed through styling to suggest incipient connections.
\end{minipage}



\vspace{1em}
\noindent
\begin{minipage}{\cardsize}
\includegraphics[width=\textwidth]{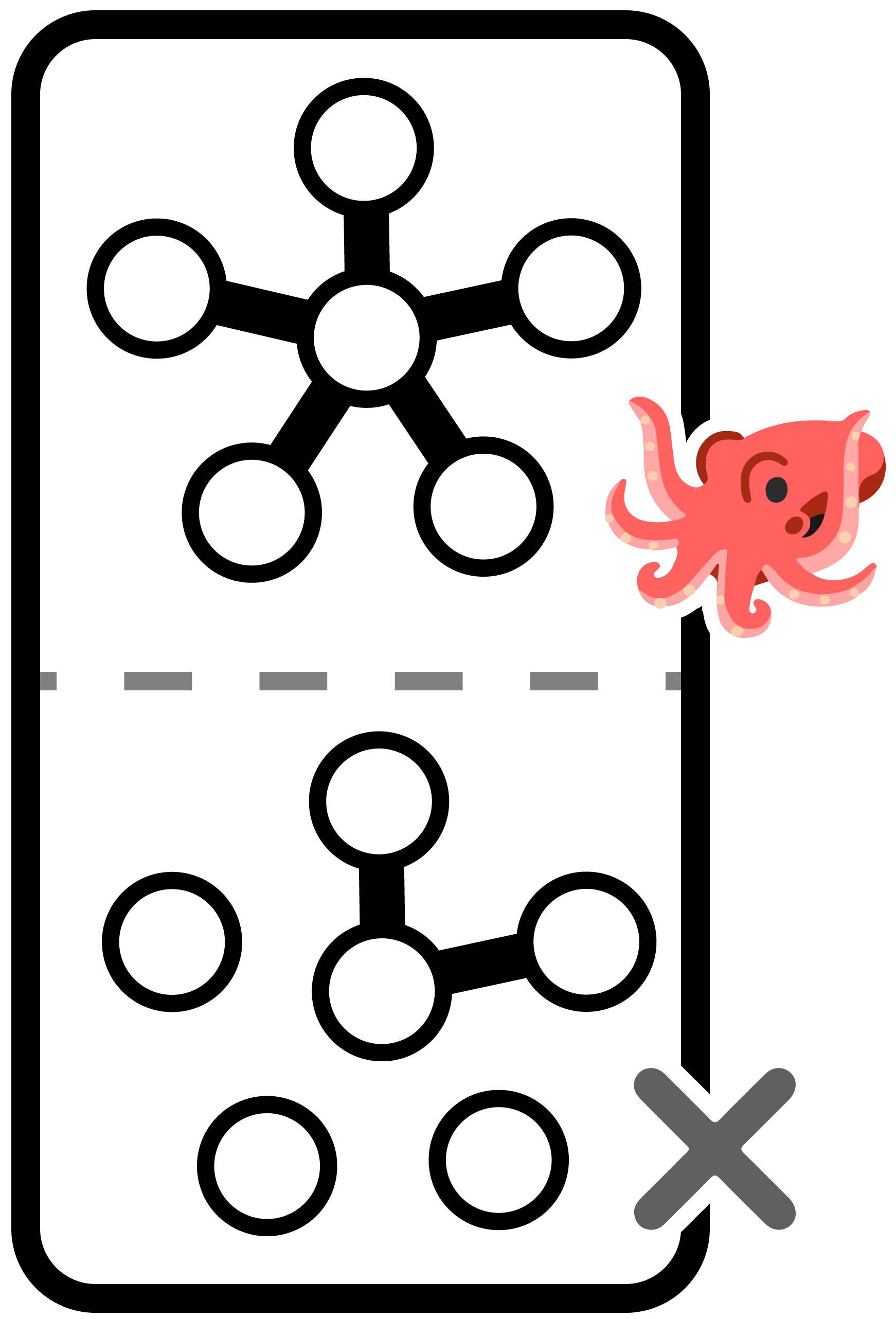}
\Description{A card with two network diagrams, the top one representing our concept of reach by showing all nodes connected to the center, with the bottom one showing only some nodes connected to the center.}
\end{minipage}
\begin{minipage}{\cardparasize}
 \noindent\textbf{Reach}: \textit{These tentacles are connected to many, most, or even all of the potential targets.} The idea that ``everything is connected'', is fundamental in conspiratorial thinking~\cite{knight2002conspiracy}. The juxtaposition of graph structure and cartographic space in octopus maps affords two ways of communicating omnipresence: through large numbers of edges, or having the edges take up large portions of the visual or cartographic space (e.g., the long reach of tentacles in \autoref{fig:england}).
\end{minipage}

    

\vspace{1em}
\noindent
\begin{minipage}{\cardsize}
\includegraphics[width=\textwidth]{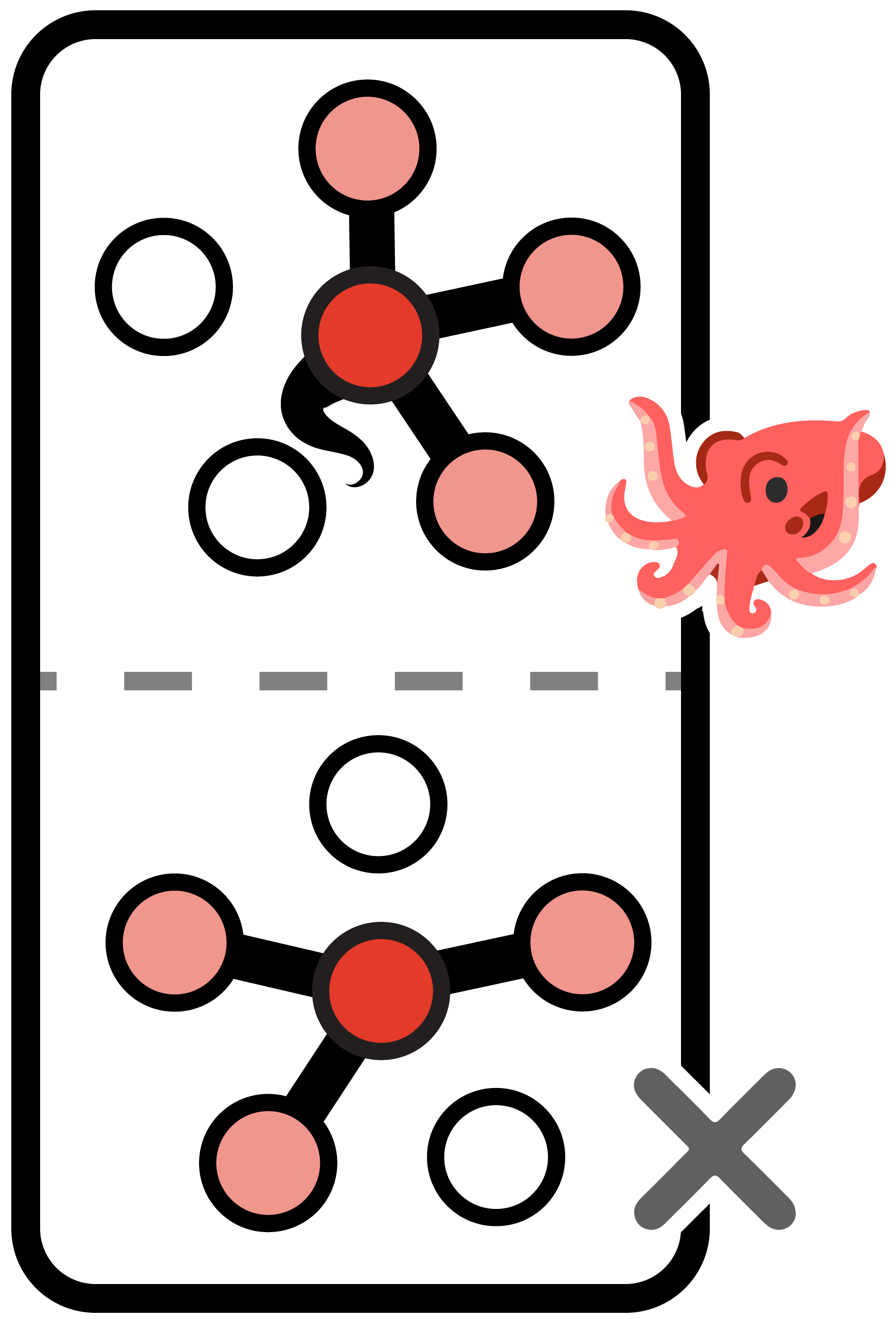}
\Description{A card with two network diagrams, the top one representing our concept of intentionality by showing a node with a wavy prospective edge reaching out towards a disconnected node, with the bottom one showing only connected and disconnected nodes with no prospect of new connections.}
\end{minipage}
\begin{minipage}{\cardparasize}
\noindent \textbf{Intentionality}: \textit{Despite the occasional independent movements of visualized threats, they are ultimately perceived to be part of an intentional strategy.} Conspiracy theories often assign agency to small groups~\cite{douglas2023whatareconspiracy, douglas2018whyconspiracy} with the ``intention to conspire''~\cite{Nera2022opportunisticpower}. Visually, intentionality can be signaled through labels, direct cues like a sequential order of targets, with tentacles grasping for new targets, or more subtle cues like the gaze behavior of the octopus (as in \autoref{fig:usa})--- even subtle choices in visualization design can produce perceptions of causality~\cite{xiong2019illusion}.
\end{minipage}


\vspace{1em}
\noindent
\begin{minipage}{\cardsize}
\includegraphics[width=\textwidth]{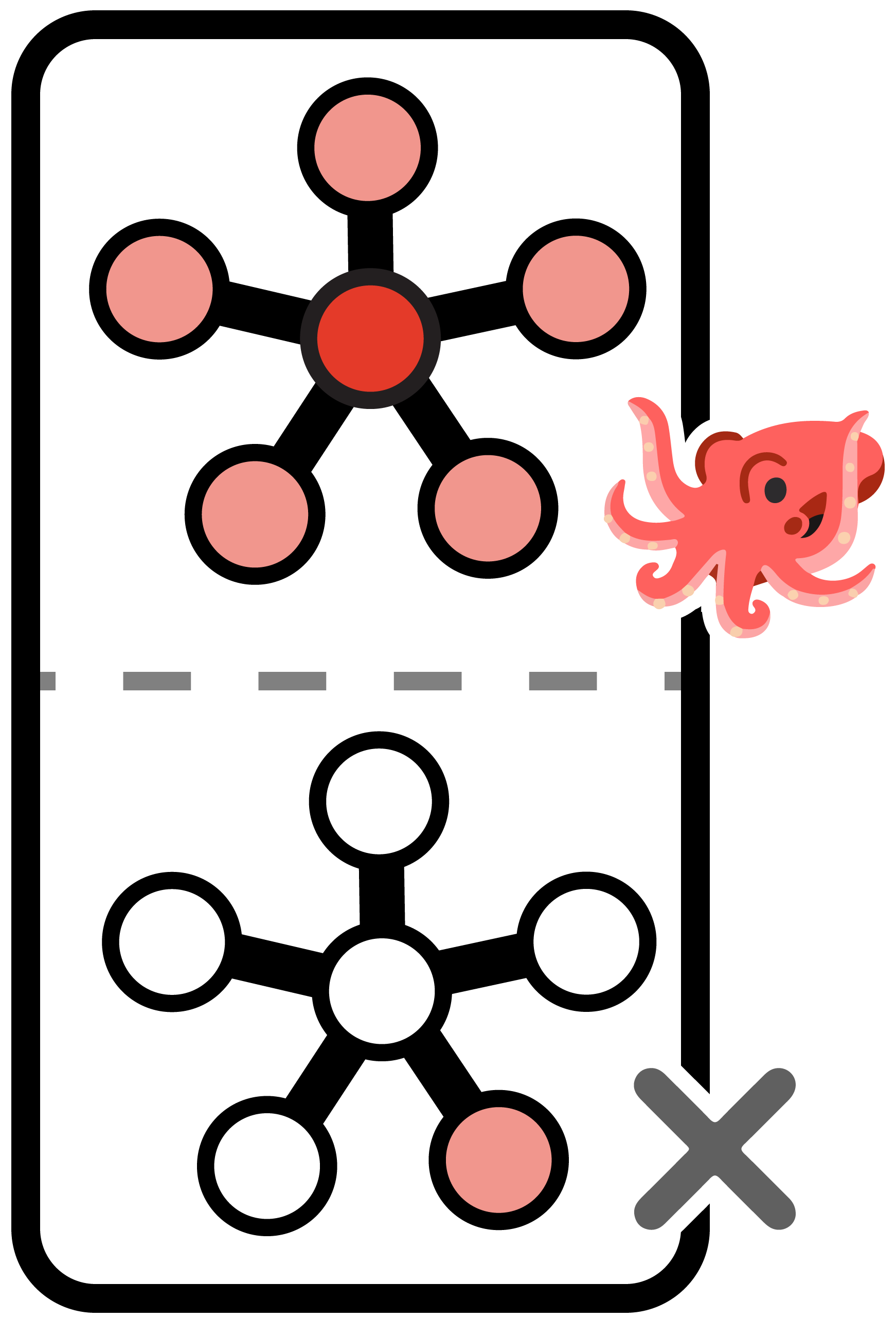}
\Description{A card with two identical network diagrams, except that in the top one a dark red node is in the center,representing our concept of grabbiness by connecting to all lighter red nodes at the edges, while in the bottom network the same configuration has only one edge node in light red, and others remain blank.}
\end{minipage}
\begin{minipage}{\cardparasize}
\noindent \textbf{Grabby}: \textit{The action of these tentacles is to acquire or control their targets. } 
    Denoting a relationship is not enough; the octopus has to have some influence over the objects in the grasp of its tentacles. Studies have linked a sense of powerlessness, often blamed on an external entity thought to have taken that agency away, to a propensity for conspiratorial thinking ~\cite{stojanov2020does,stojanov2021examining,douglas2018whyconspiracy,imhoff2014speaking}. This grabbiness can be very literal, with arms wrapping around victims (\autoref{fig:benevolent}), or it can pierce through the target, as shown in the NATO octopus~\cite{Zanin2021Octopus}. Mere connection to a large number of targets could instead denote reciprocal or mutual relationships.
\end{minipage}

\vspace{1em}
\noindent
\begin{minipage}{\cardsize}
\includegraphics[width=\textwidth]{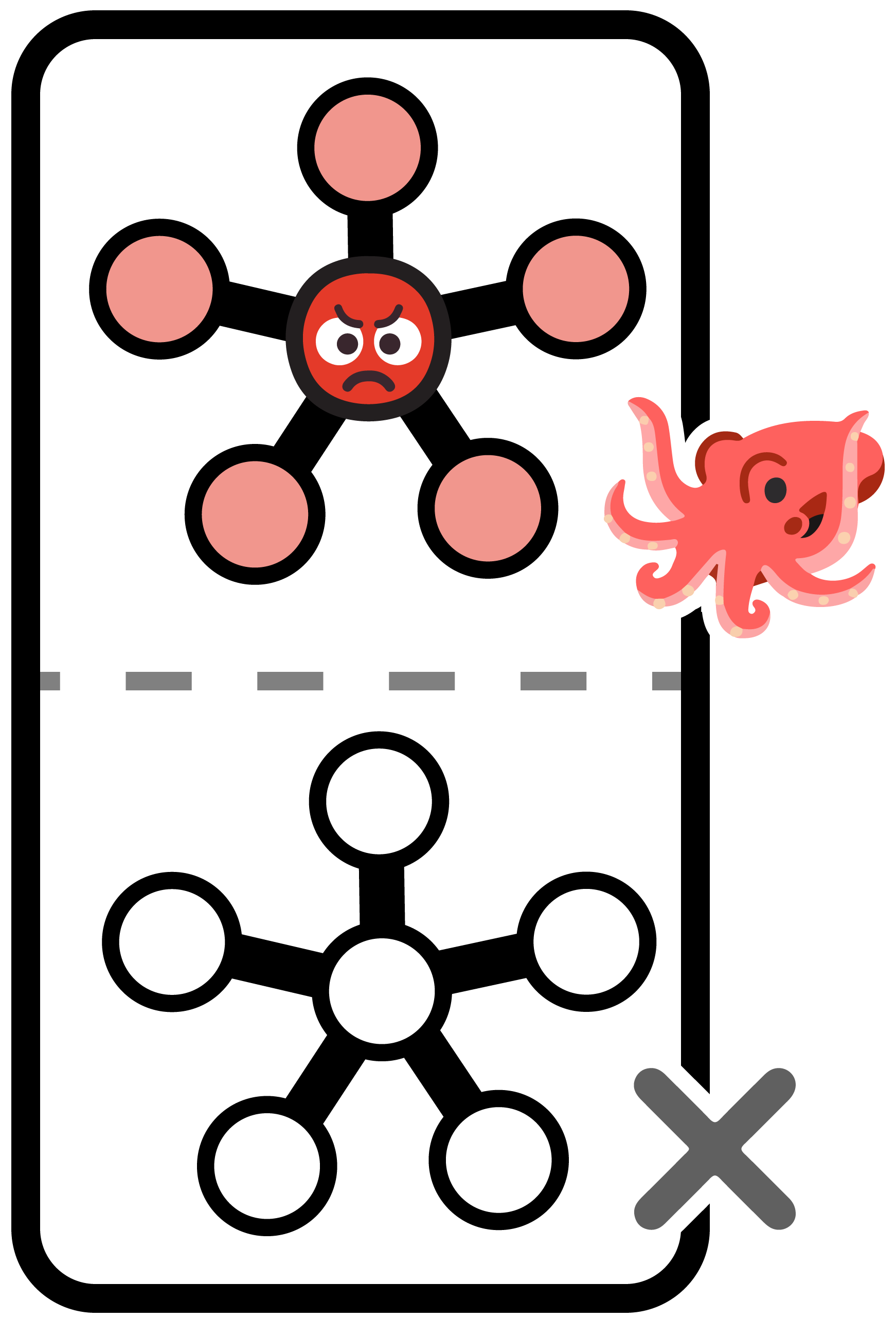}
\Description{A card with two identical network diagrams, the top one representing our concept of threat through both the use of red nodes and a set of menacing eyes on the central node, while the bottom figure excludes both color and eyes.}
\end{minipage}
\begin{minipage}{\cardparasize}
\noindent \textbf{Threat}: \textit{The ultimate goal of this octopus is, from the perspective of the reader, nefarious or threatening.} While benevolent octopus maps do exist (see \autoref{fig:benevolent}),  in most cases conspiratorial thinking requires ``malevolent or forbidden acts'' \cite{douglas2023whatareconspiracy}. Threat is difficult to convey through data alone, but relies on framing effects and assumptions. Visually, this threat can be indicated through the expression of the octopus (although we note a variety of expressions in our corpus, from the malevolent glee in \autoref{fig:subway} to the almost bashful Austro-Hungarian octopus in \autoref{fig:prussia}) or portraying negative impacts of the tentacle connections (such as targets struggling under a tentacle's grasp). However, non-visual elements like biased or slanted titles can also influence the perception of patterns in data~\cite{kong2018frames}.
\end{minipage}

\vspace{1em}
Taken together, these interconnected components present an implicit visual argument: a centralized, nefarious body is employing multiple levers of control in multiple places to undermine or victimize large parts of some region or system.
There are occasional modifications or exceptions to this general form.
For instance, drawing a connection from an octopus to the semiotically similar gorgon or hydra~\cite{schnier1956morphology}, one can depict one or more tentacles being severed, either to show that the threat will remain as long as the central body remains, and/or that the task of defeating the octopus may be long and arduous. Centrality can also be violated, such as Austria-Hungary acting as a second octopus under Prussia's implied control in \autoref{fig:prussia}. In  \S\ref{sec:edges} we also note the existence of the \textit{benevolent} octopus, where, rather than threatening, the octopus figure is meant to be a figure of self-identification or even exterior protection.




\section{Evaluating the Octopus Map}
\label{sec:experiment}
We performed a preregistered\footnote{\url{https://osf.io/8n5ah/?view_only=7f14c46cd644467fbe89c6ce4c01eb2a}} crowd-sourced study to assess the \textit{rhetorical impact} of the \textit{visual} and \textit{structural} components of the octopus map. That is, while the designers of such maps may \textit{intend} for the octopus to be read as a centralized nefarious force, is this how these maps are \textit{interpreted} by mass audiences? And would a less sensationalized version of the same information accomplish the same rhetorical goals? Of particular interest to us was whether, even for maps without explicitly octopus-inspired motifs, some of the more ``implicit'' signatures of octopus maps could inspire similar conspiratorial or adversarial responses. Motivating this question was prior work suggesting that these techniques (like directional arrows and threatening colors) can produce xenophobic or threatening perceptions of migrant data or other sorts of population flows~\cite{van2020migration}, as well as the existence of, rather than a simple dichotomy between sensationalized and non-sensationalized persuasive maps, a continuum of attitudes around how the look and feel of maps are seen as sensationalized~\cite{muehlenhaus2012if}.

Our intention with this study was therefore neither to fully map out the design space of octopus maps nor to precisely quantify how persuasive these maps are. The enormous visual and dataset variability of the maps just as encountered in our corpus (let alone in the space of persuasive cartography more generically) would seem to preclude a thorough investigation of the full combinatorial design space, especially for a research team of our size. The key importance (and sensitivity) of framing information~\cite{hullman2011rhetoric} and individual priors~\cite{pandey2014persuasive} in visualization rhetoric would also limit the generalizability of any quantitative results on persuasion. Rather, our goal was to assess the ability of the central argument of the octopus map to ``survive'' across forms that lack one or more of the visual or data-based traits we identify in \S\ref{sec:parts}.

We wished to provide an interpretative task that afforded both positive and negative interpretations of the data. We therefore opted to assess participant responses to octopus and octopus-like maps of a fictitious militarist country in a fictitious geographic region. This scenario limits our ability to draw tight connections with how sensationalist framings of data connect to real-world shifts in opinion or propensity to action, but allows us to more directly test for how different features of the data or design of these maps produce rhetorical impacts, without having to collect detailed information on prior attitudes (as in similar empirical looks at persuasive visualizations such as Pandey et al.~\cite{pandey2014persuasive}) or bring up as many of the strong emotions associated with these maps in real-world settings. Our decision was motivated by the entanglement of these maps with existing real-world sources of hate and discrimination, and general sensitivity around the subjects of the octopus maps in our corpus. We also note that prior work has shown that elaborate narrative framings do not always have predictable or expected impacts on crowd-sourced task performance~\cite{dimara2017narratives}. We therefore elected for a relatively simple framing: a fictitious country has a number of military bases in neighboring countries in the region. While this scenario is by no means free from connections and prior beliefs about, e.g., the military in general (and we include an exploratory analysis searching for these real-world connections), it was our supposition that the emotional affect engendered by the manipulation of the various octopus-like features of these data would have rhetorical impact--- that participants would see the country as, e.g., engaged in a plot for regional hegemony, or engaged to undermine the sovereignty of its neighbors, or even as a benevolent protector. 

\textbf{Additional study information including survey instruments, stimuli, anonymized participant data, and analyses, are available at \url{https://osf.io/56e9u/}.
}

\subsection{Methods}
After soliciting consent, we showed participants an octopus or octopus-like map of a highlighted country within a fictitious geographic region with the following prompt:
``This is a map showing a fictional country, Huskiland, that appeared in an international newspaper. The map depicts Huskiland’s military bases in the region.'' We chose this framing as it affords both benevolent (Huskiland as leader of a protective military alliance) and malevolent (Huskiland as regional threat) interpretations.

Participants then rated their agreement with the following statements on a 7-point rating scale from 1-Strongly disagree to 7-Strongly agree. Each of these questions was intended to directly link to the components of the visual argument of the octopus map as discussed in \S\ref{sec:parts}. The label prefixing each question was not presented to participants, but denotes which question corresponds to which rhetorical component:

\begin{enumerate}
    \item \textbf{Centrality}: Huskiland is a central military power in the region.
    \item \textbf{Tentacularity}: Huskiland is expanding its military reach.
    \item \textbf{Reach}: Huskiland is already present in many of the countries in the region.
    \item \textbf{Intentionality}: Huskiland's placement of bases is part of an intentional military strategy.
    \item \textbf{Grabby}: Huskiland uses these bases to exert military or political control over its neighbors.
    \item \textbf{Threat}: Huskiland is a threat to the peace and stability of the region.
\end{enumerate}

Note that these questions are rhetorically loaded: the data itself did not provide direct evidence for or against many of these propositions. Rather, our goal was to solicit how much of the implicit or explicit visual argument was communicated in the resulting map. As a first order attempt to capture the success or failure of the ``octopus map'' visual argument, we sum the scores of these six questions into a single ``octopodality'' scale value; many of our subsequent analyses are based on this overall value, as a proxy for the overall ``success'' of the visual argument (although note that, commensurate with other scales developed as part of psychological research, we would not expect, nor did we empirically observe, individual question components of this scale to have identical distributions or variability; see \S\ref{sec:exploratory} for more details).

In addition to a CAPTCHA on the survey itself, we included a validation question as a check for comprehension of the underlying map data and to exclude ``click-through'' behavior. Participants who did not correctly answer this question were compensated but otherwise excluded from our analysis:
\begin{itemize}
    \item \textbf{In how many countries in the region does Huskiland have military bases?}
\end{itemize}

We lastly solicited a free-text question:
\begin{itemize}
    \item \textbf{How would you describe the relationship between Huskiland and its neighbors in the region?}
\end{itemize}

After the main tasks, we collected participant demographics. We considered including post-tests for assessing both visualization literacy (e.g., the VLAT or mini-VLAT~\cite{minivlat}) and propensity for belief in conspiracies~\cite{brotherton2013measuring}, but ultimately elected not to collect additional information beyond gender, age, education, and self-reported familiarity with maps and charts. This decision was driven by both a desire to keep the task as short as possible, and also a lack of strong hypotheses linking visual or data features to high or low scores on these scales. This lack of solicitation of prior attitudes is somewhat at odds with prior work on persuasive visualizations~\cite{pandey2014persuasive}, but we note that a) our decision to employ fictional settings was meant to reduce (although not eliminate) the impact of potential biases or prior beliefs and b) our research questions (around the \textit{preservation} of the general visual argument of the octopus map across less explicitly sensationalized forms) are less concerned with the extent to which the octopus map's visual argument is ultimately \textit{successful} in, e.g., changing people's minds about the potential malevolence of a particular entity.

\subsubsection{Stimuli}
Each stimulus was a variation of the same base map of a fictional continent. On this continent, Huskiland was a centrally located country connected to several other countries in the region. To make the map itself, we used a Fantasy Map Generator \cite{haniyeu2017fantasymap}, to weaken any potential connection between our maps and real-world geopolitics.

The primary variation among stimuli was the representation of Huskiland and its bases in neighboring countries. In line with Muehlenhaus'~\cite{muehlenhaus2013design} observation that different cartographic designs are viewed as more or less persuasive or sensationalized, we wished to explore a gradient of varying map designs, ranging from the relatively straightforward maps that avoid the menacing arrows and sensationalized colors of ``propaganda cartography''~\cite{van2020migration} all the way to what we viewed as the canonical octopus form (see \autoref{sec:canon}). We note that this space of potential map designs is quite large. In general terms, there are many representatives along this potential gradient of seemingly neutral to fully sensationalized maps, each of which with their own design parameters around the use of color, map projection, glyph design, titling, etc. Even within the specific condition of the octopus map, we note diversity in factors such as the octopus' location, expression, style of tentacle (whether touching, grasping, or piercing its victims). We are skeptical that a single experiment could meaningfully or reliably explore more than a fraction of this space. We instead focus on this purported gradient of more or less sensationalized techniques for mapping inter-country relationships, with the goal of assessing the extent to which the central octopus argument we lay out in \S\ref{sec:parts} is or is not conveyed across these increasingly ``octopodal'' levels.

We therefore varied three factors independently across maps:

\textbf{Color}, 2 levels: Either a more neutral \textbf{gray} color, or a \textbf{red} color we intended to be more sensationalized~\cite{monmonier1996how}, sampled from a map (of a communist octopus) from our corpus.

\textbf{Edge Type}, 4 levels: We represented a connection between Huskiland and a country through the use of \textbf{color} alone, as a \textbf{node-link} diagram without directionality, as a more sensationalized form of node-link diagram with directional \textit{arrows}~\cite{van2020migration}, or using a literal \textit{octopus} motif with tentacles for edges.

\textbf{Connection}, 2 levels: Out of the 14 possible countries, Huskiland had bases in either four (\textbf{low} connection), or eight (\textbf{high} connection).

\begin{figure*}
    
\begin{center}
\begin{tabular}{cccc} 
    \adjustbox{valign=m}{\includegraphics[width=0.22\textwidth]{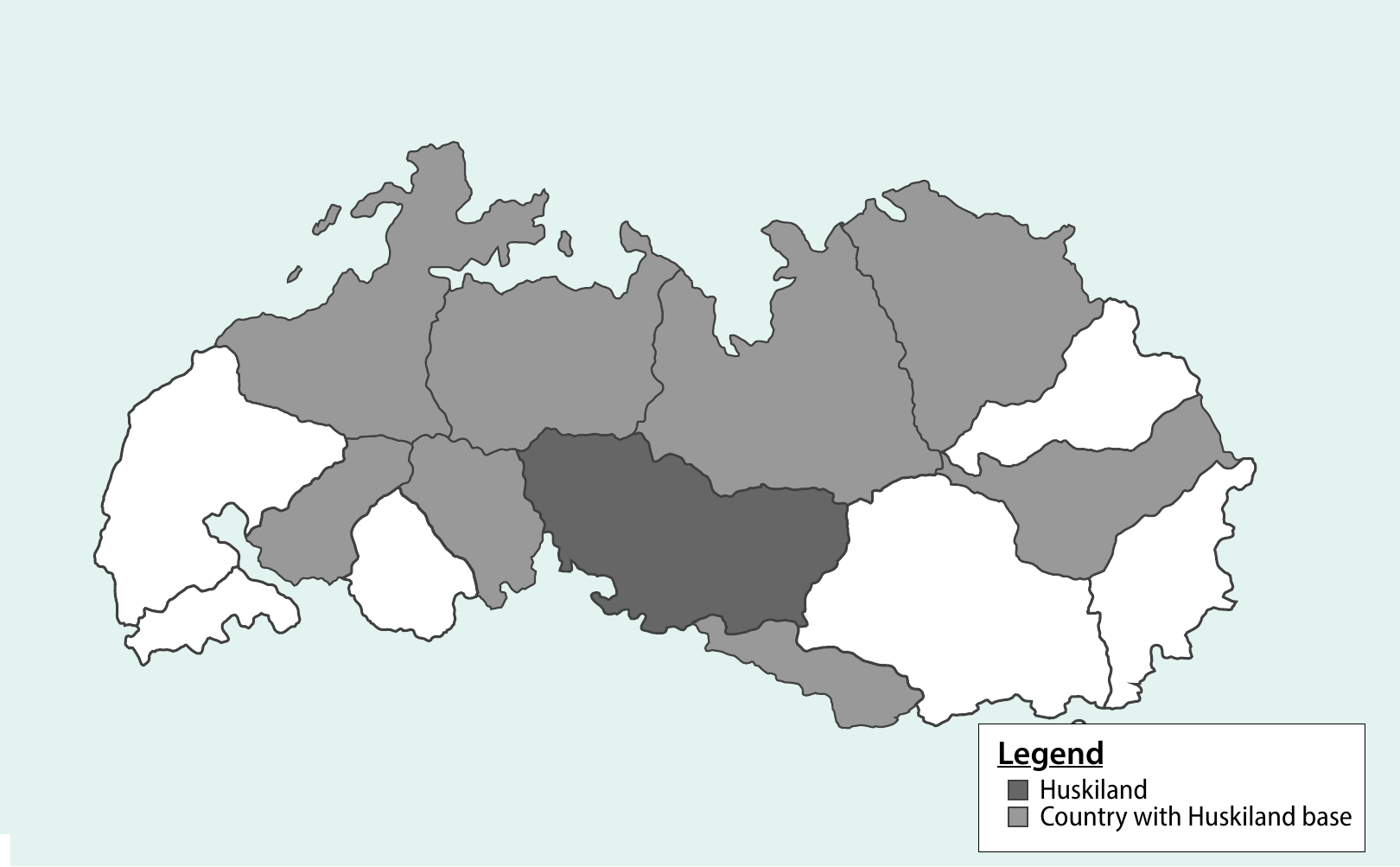}} & 
    \adjustbox{valign=m}{\includegraphics[width=0.22\textwidth]{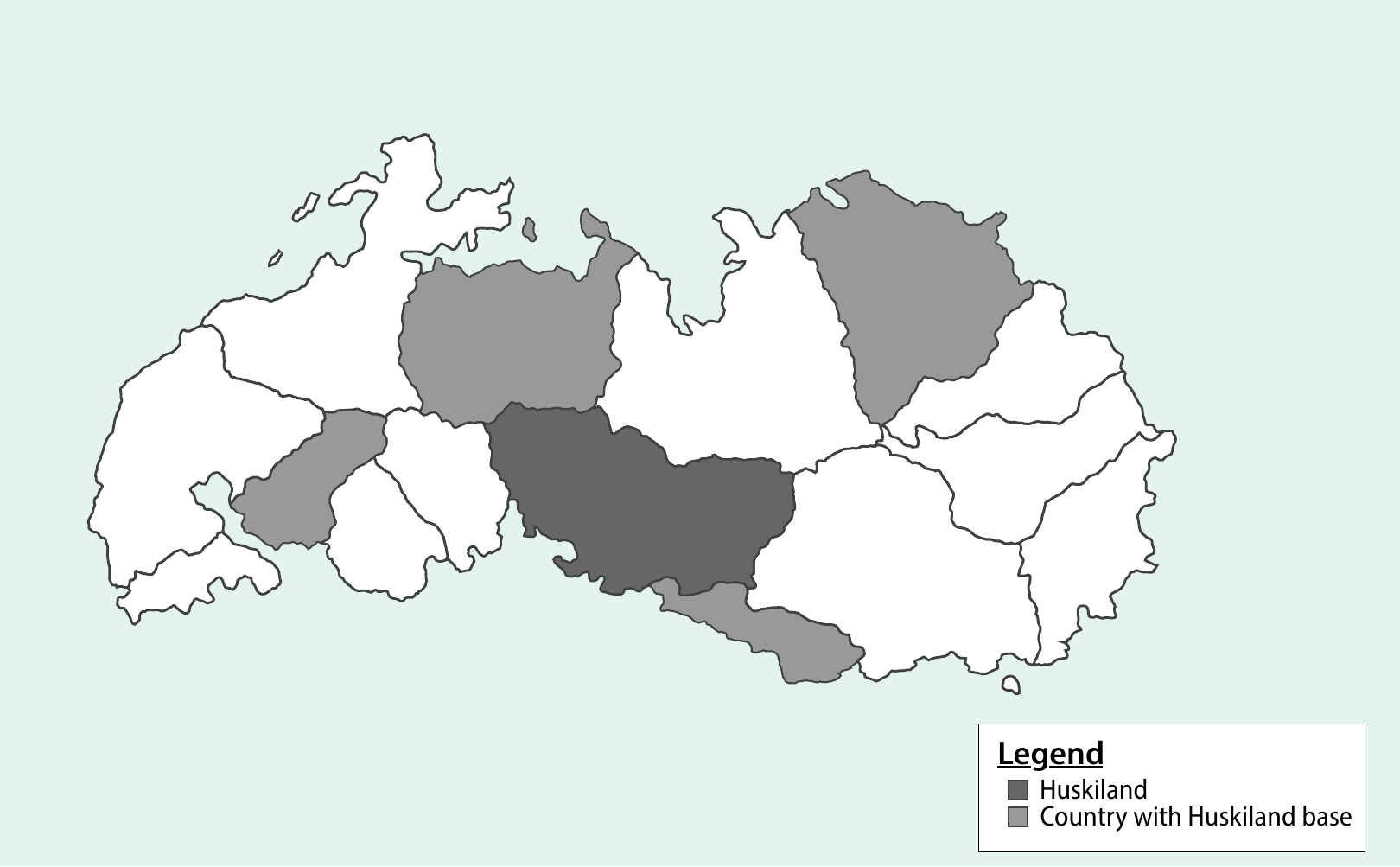}} & 
    \adjustbox{valign=m}{\includegraphics[width=0.22\textwidth]{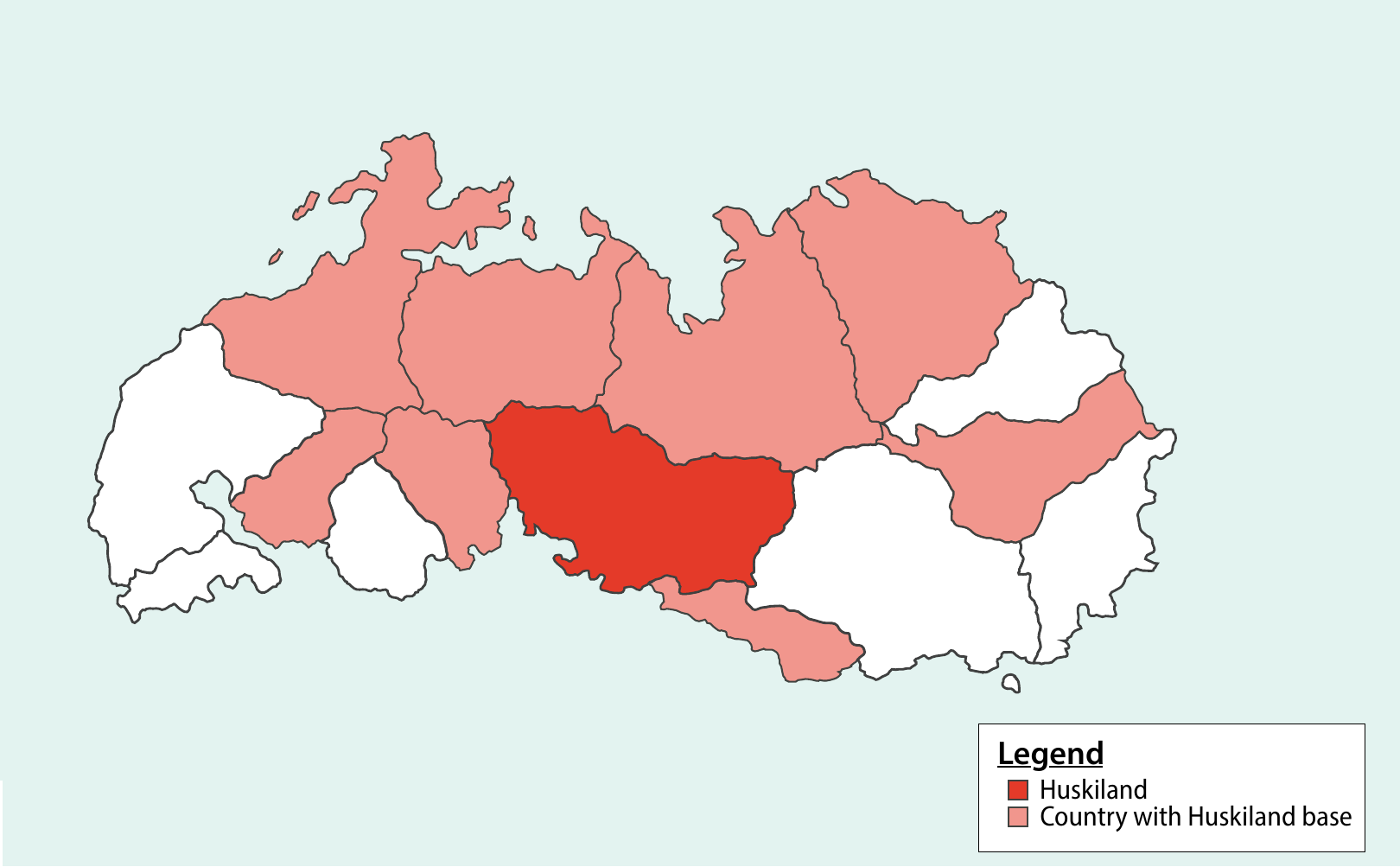}} & 
    \adjustbox{valign=m}{\includegraphics[width=0.22\textwidth]{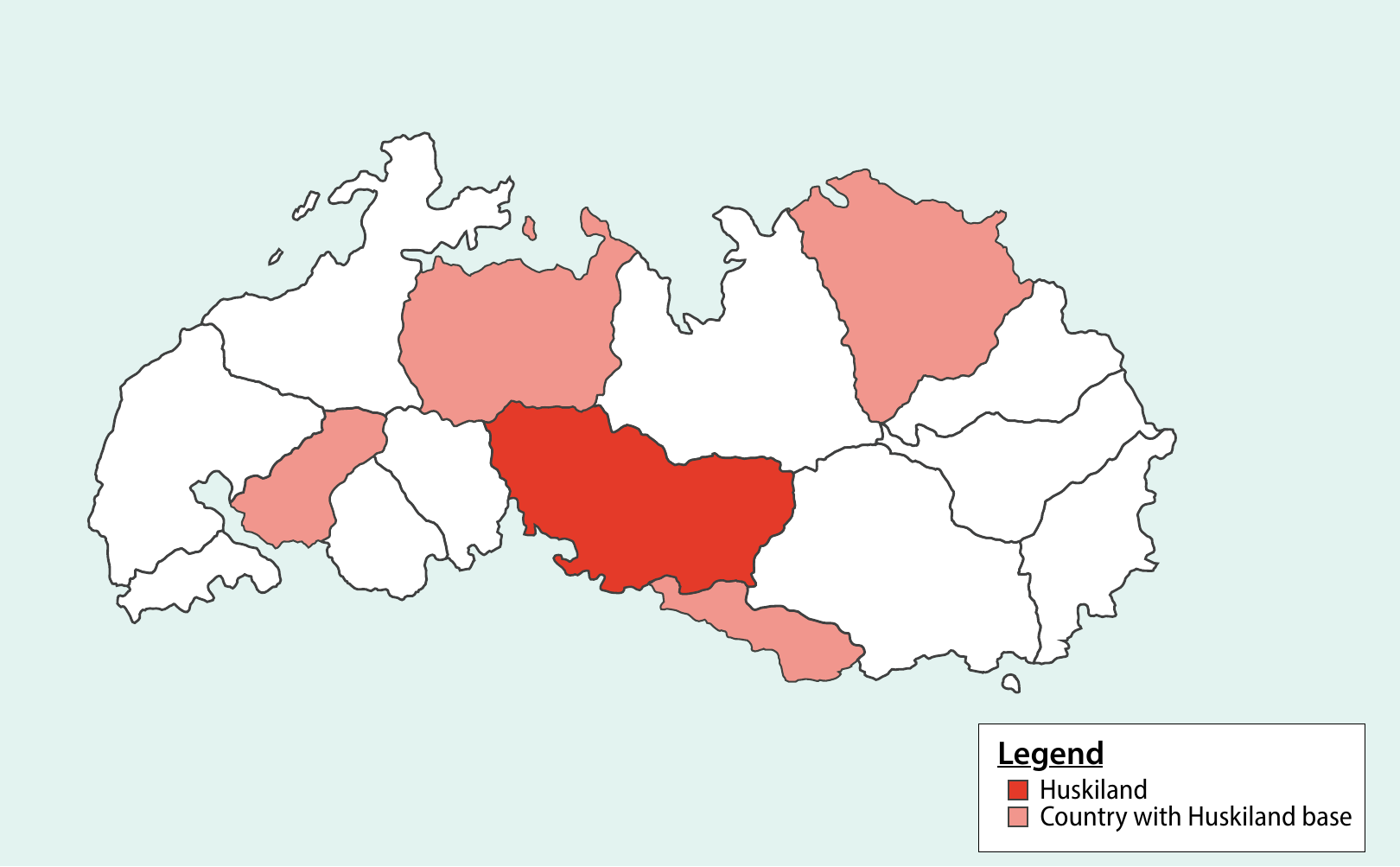}} \\
    
        \adjustbox{valign=m}{\includegraphics[width=0.22\textwidth]{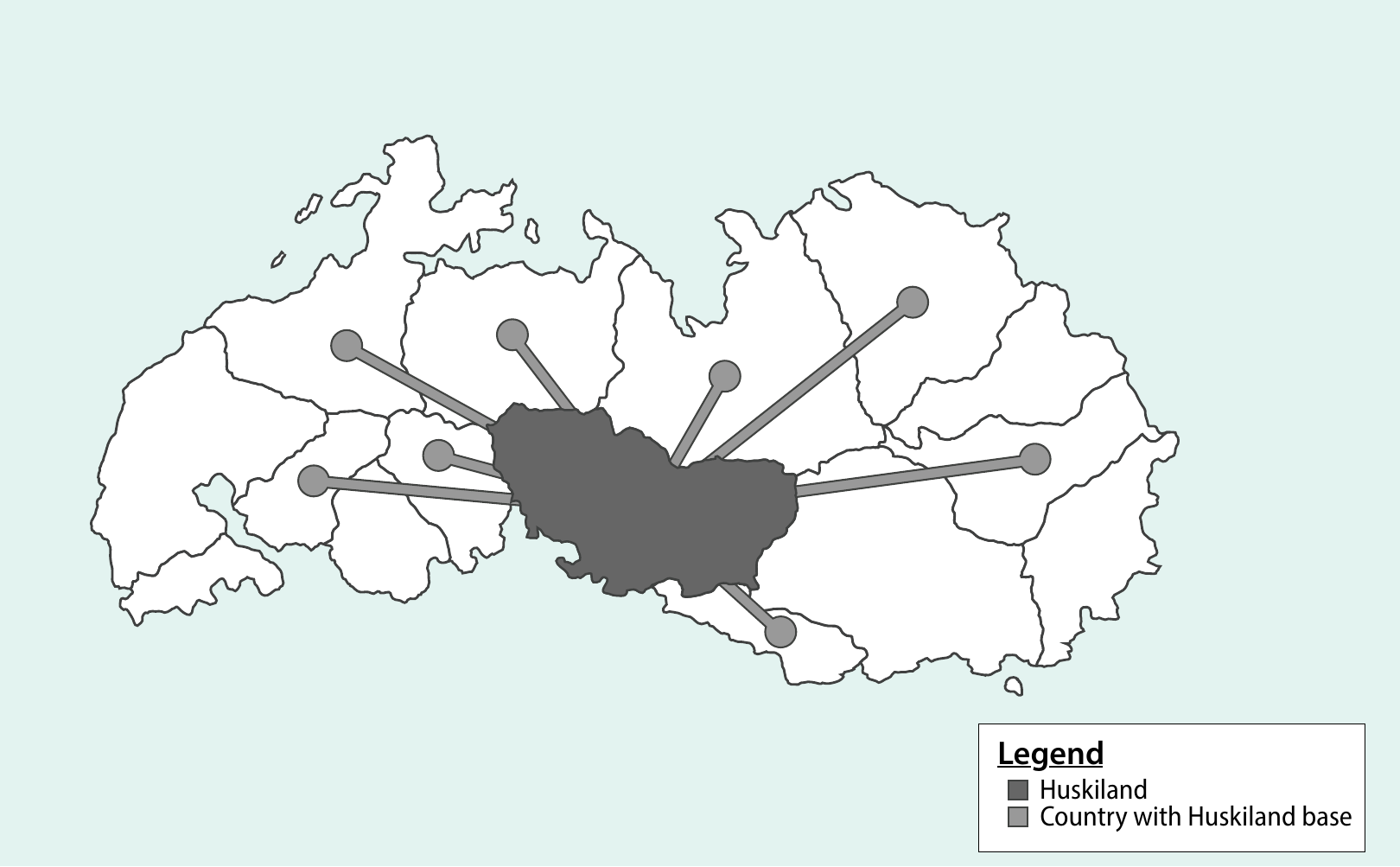}} & 
    \adjustbox{valign=m}{\includegraphics[width=0.22\textwidth]{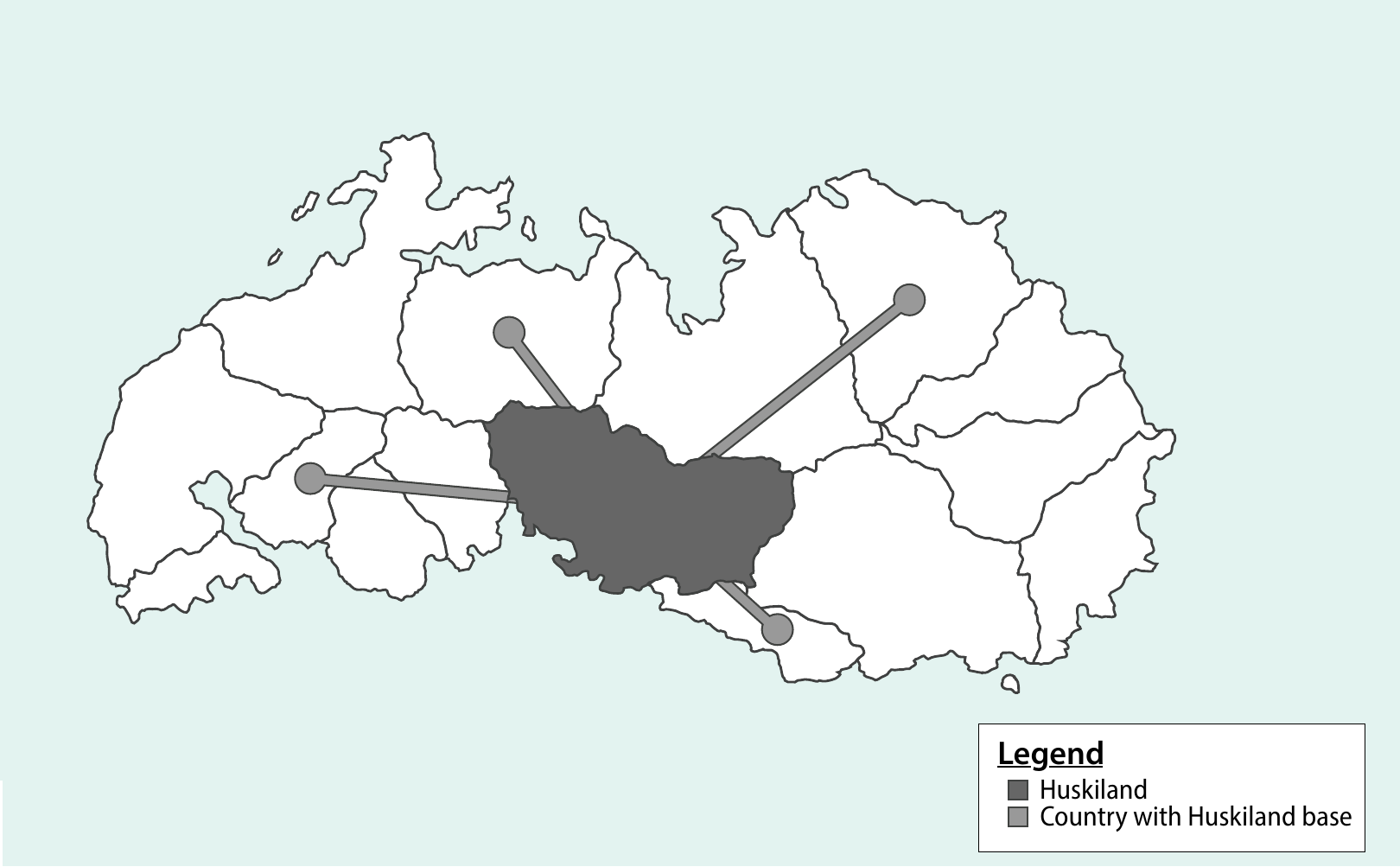}} & 
    \adjustbox{valign=m}{\includegraphics[width=0.22\textwidth]{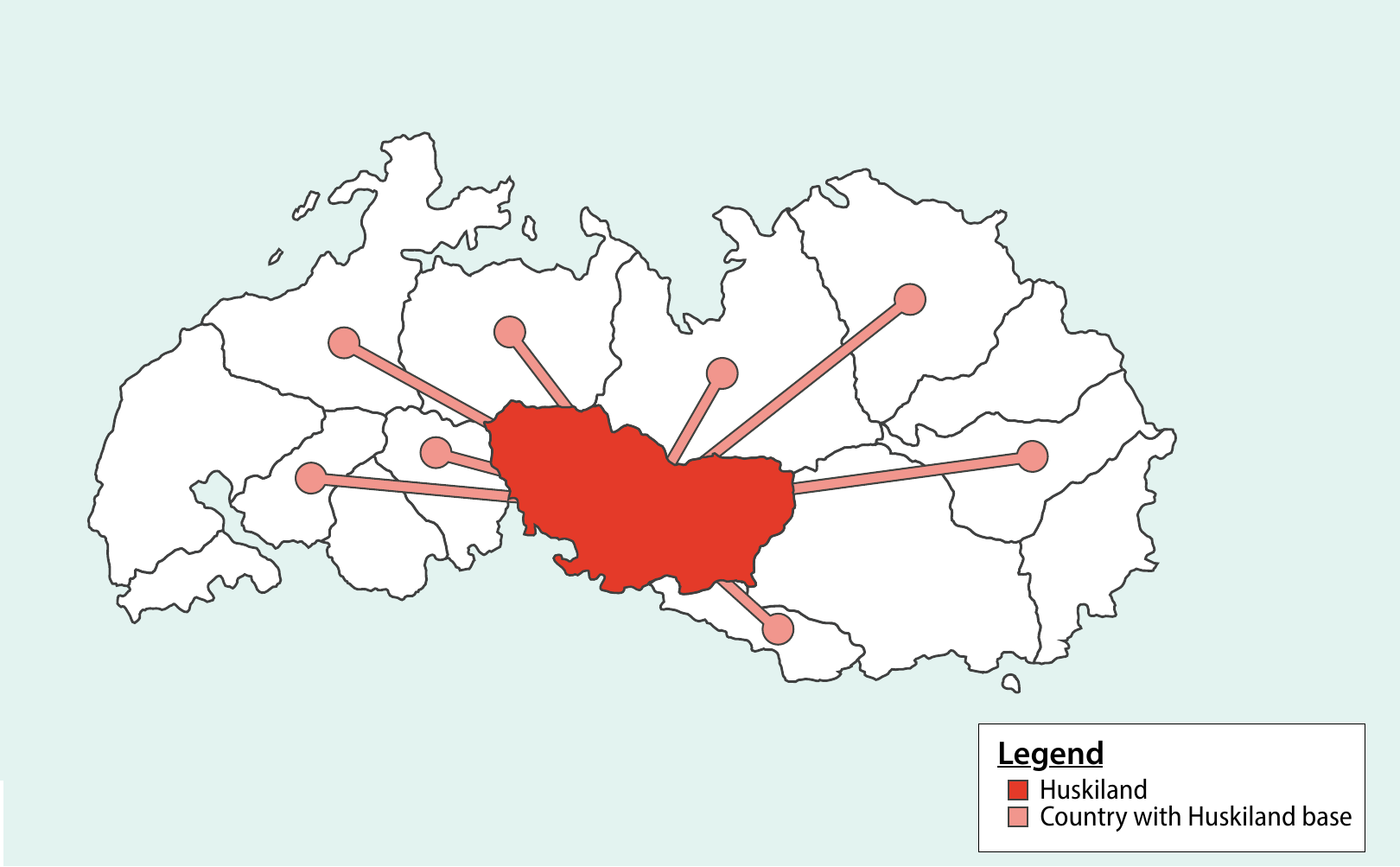}} & 
    \adjustbox{valign=m}{\includegraphics[width=0.22\textwidth]{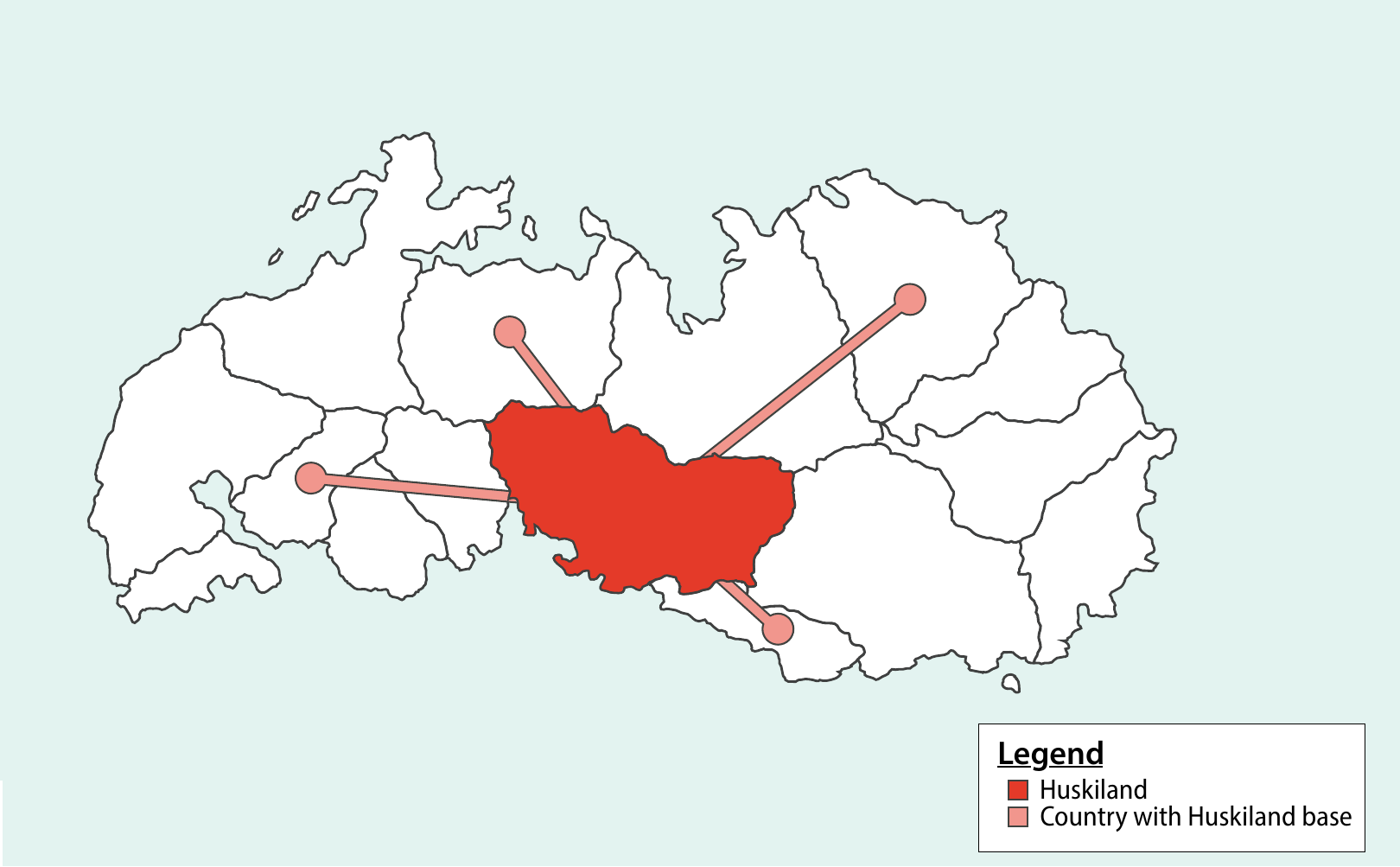}} \\

    \adjustbox{valign=m}{\includegraphics[width=0.22\textwidth]{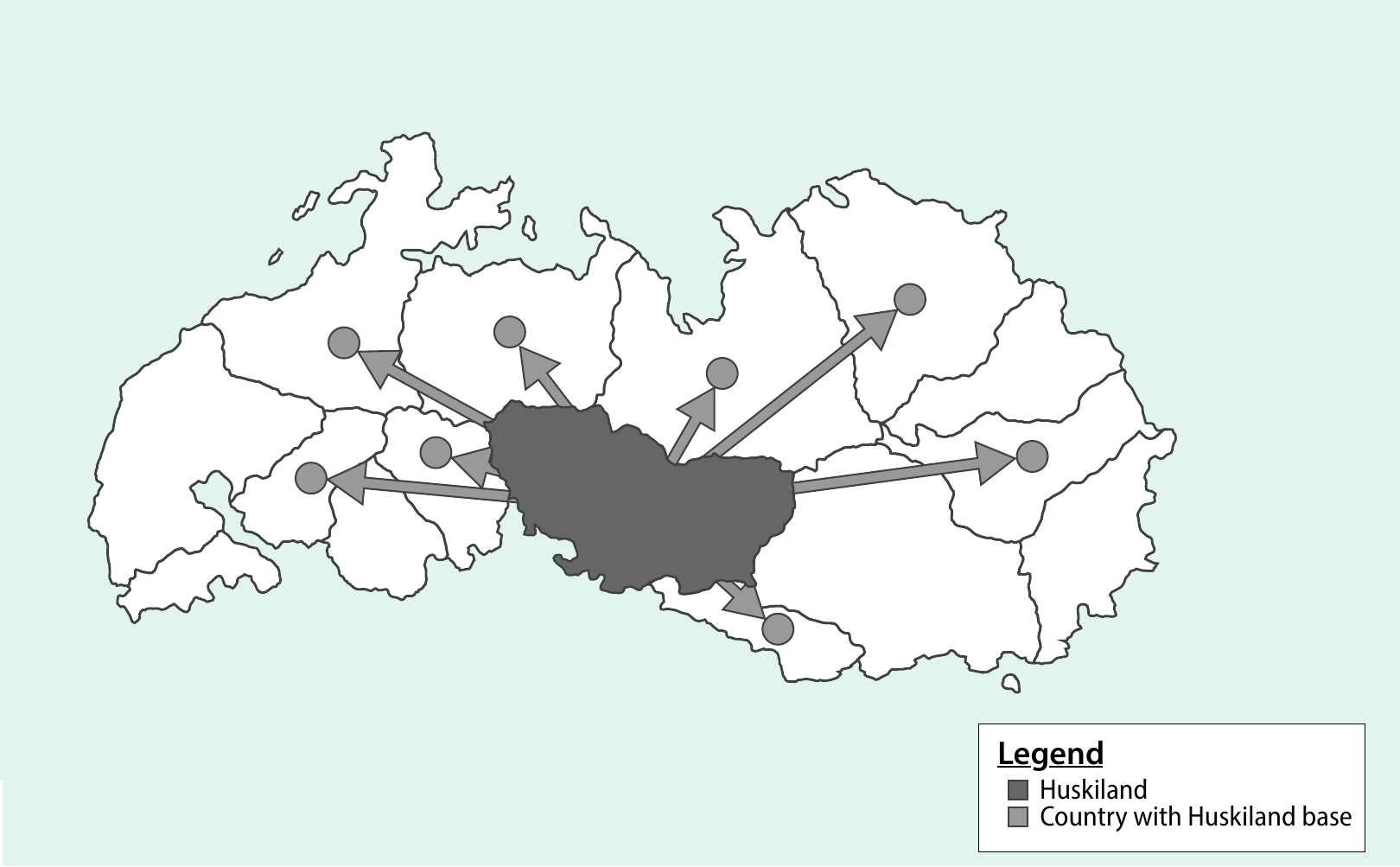}} & 
    \adjustbox{valign=m}{\includegraphics[width=0.22\textwidth]{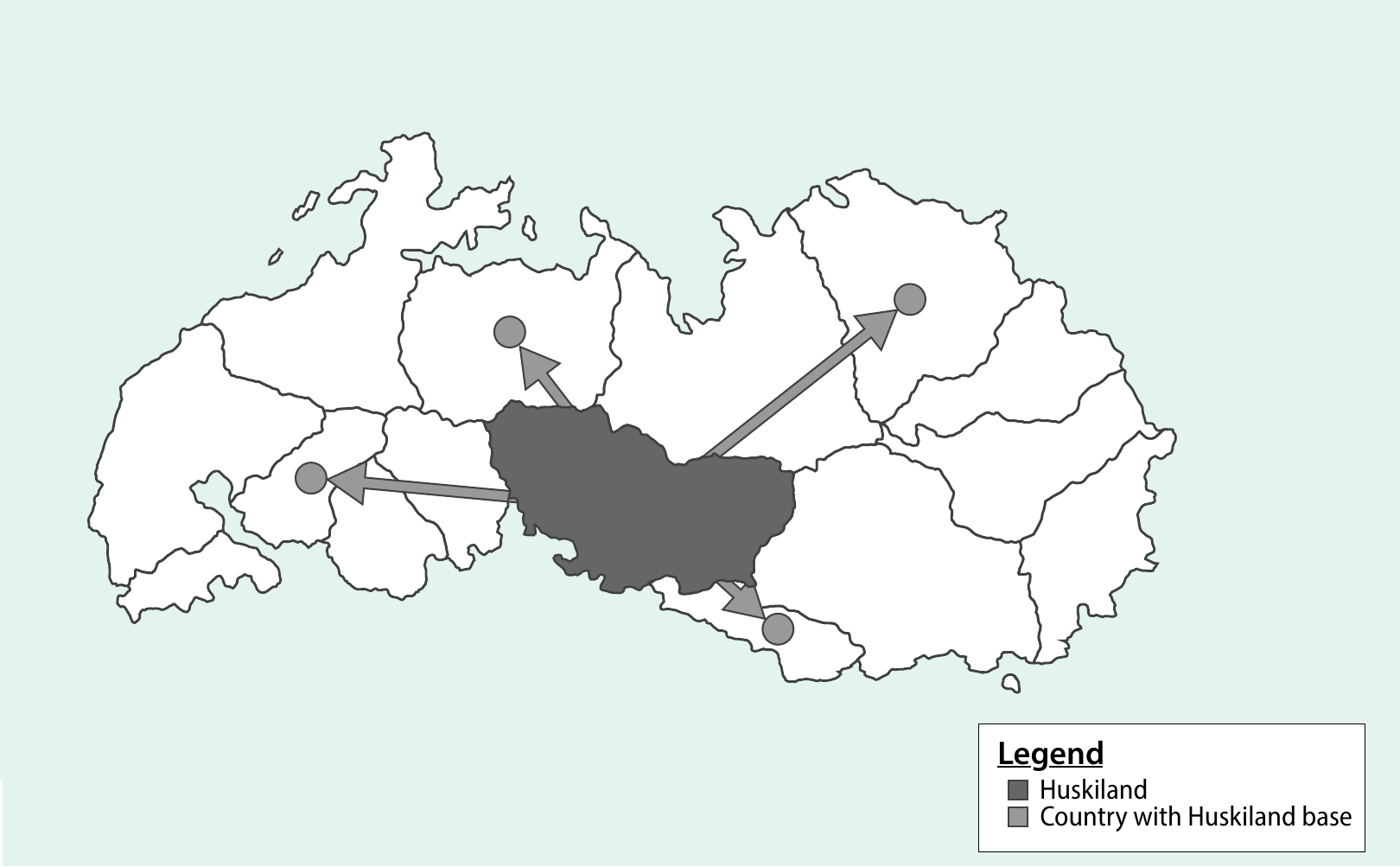}} & 
    \adjustbox{valign=m}{\includegraphics[width=0.22\textwidth]{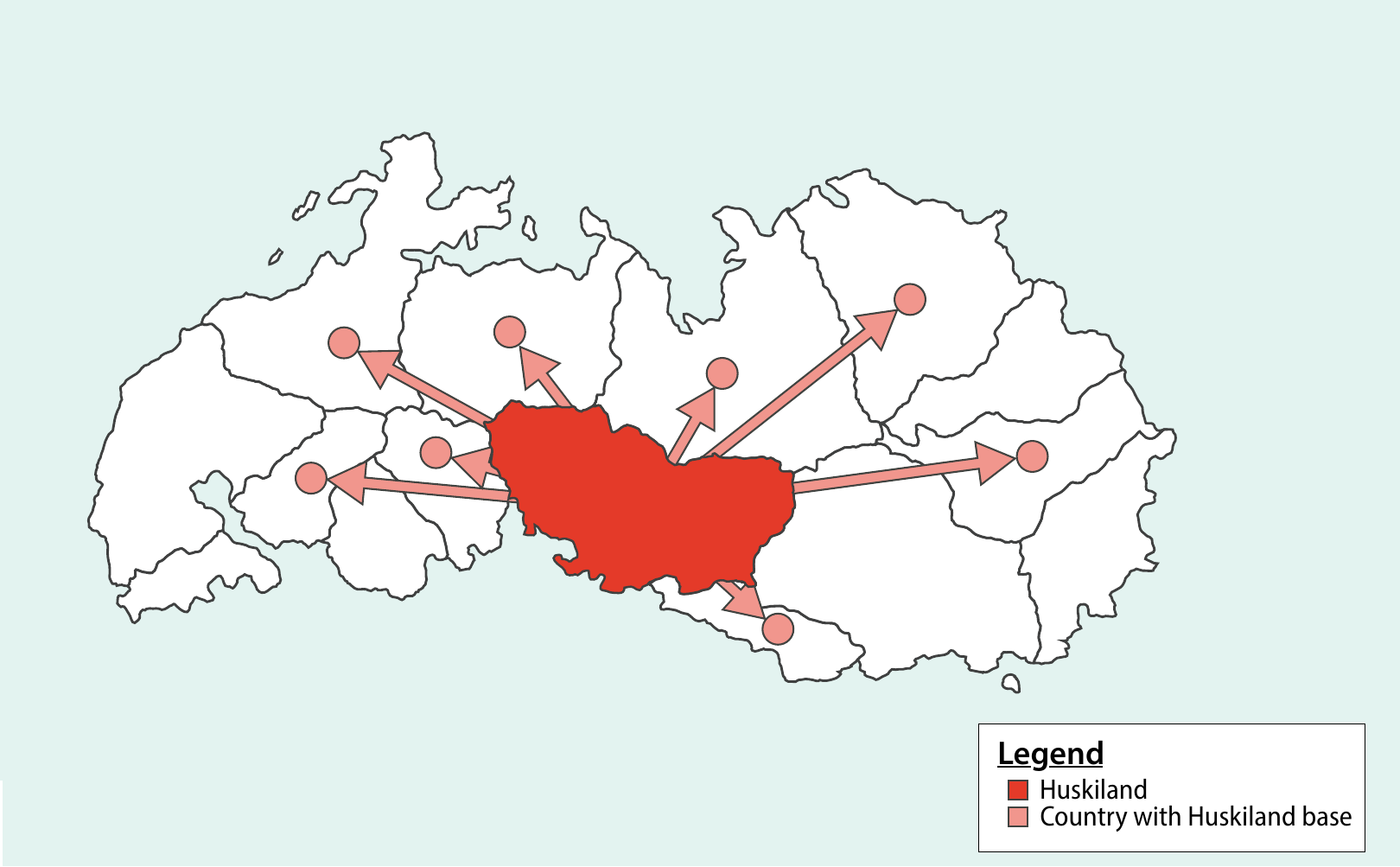}} & 
    \adjustbox{valign=m}{\includegraphics[width=0.22\textwidth]{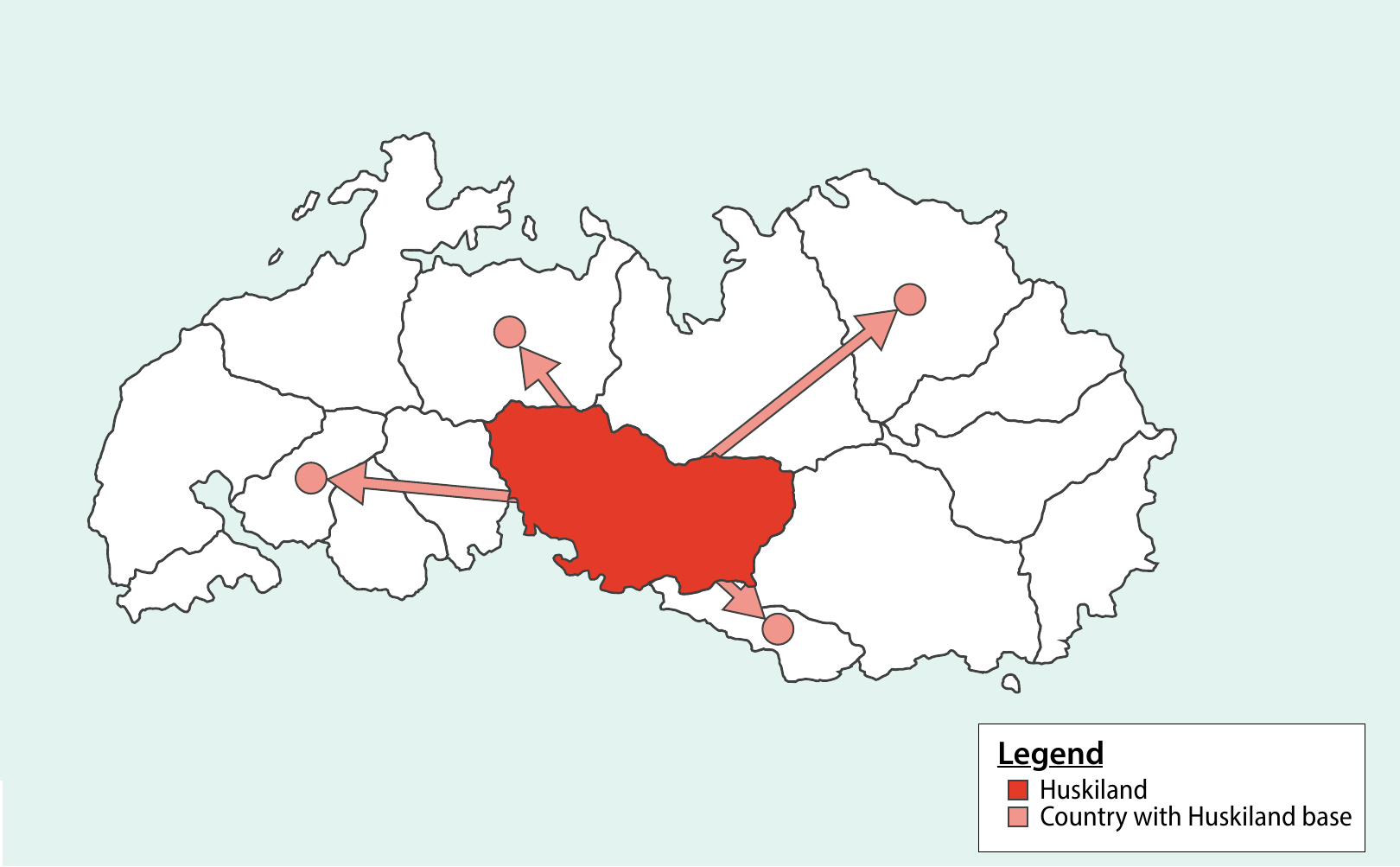}} \\

        \adjustbox{valign=m}{\includegraphics[width=0.22\textwidth]{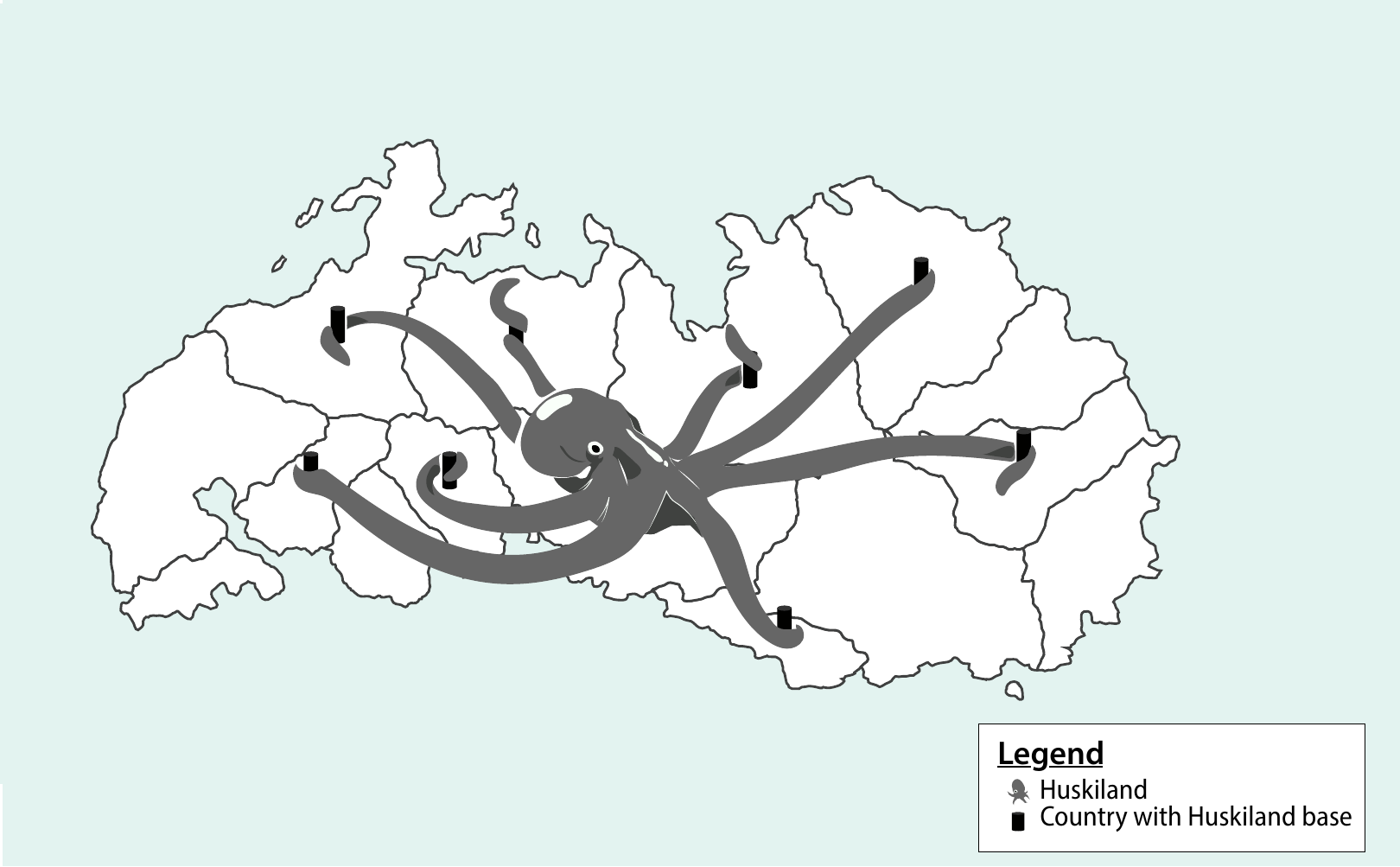}} & 
    \adjustbox{valign=m}{\includegraphics[width=0.22\textwidth]{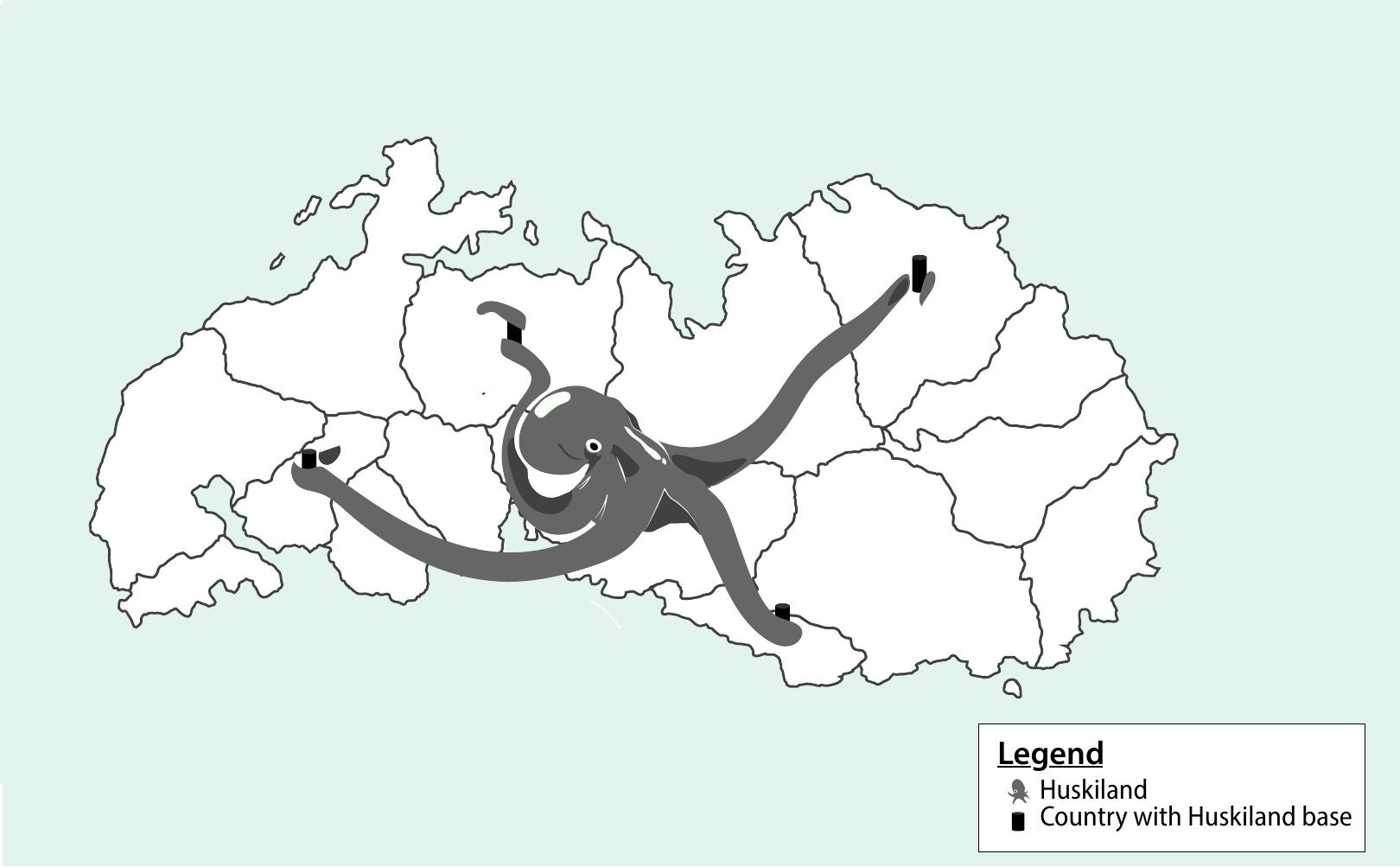}} & 
    \adjustbox{valign=m}{\includegraphics[width=0.22\textwidth]{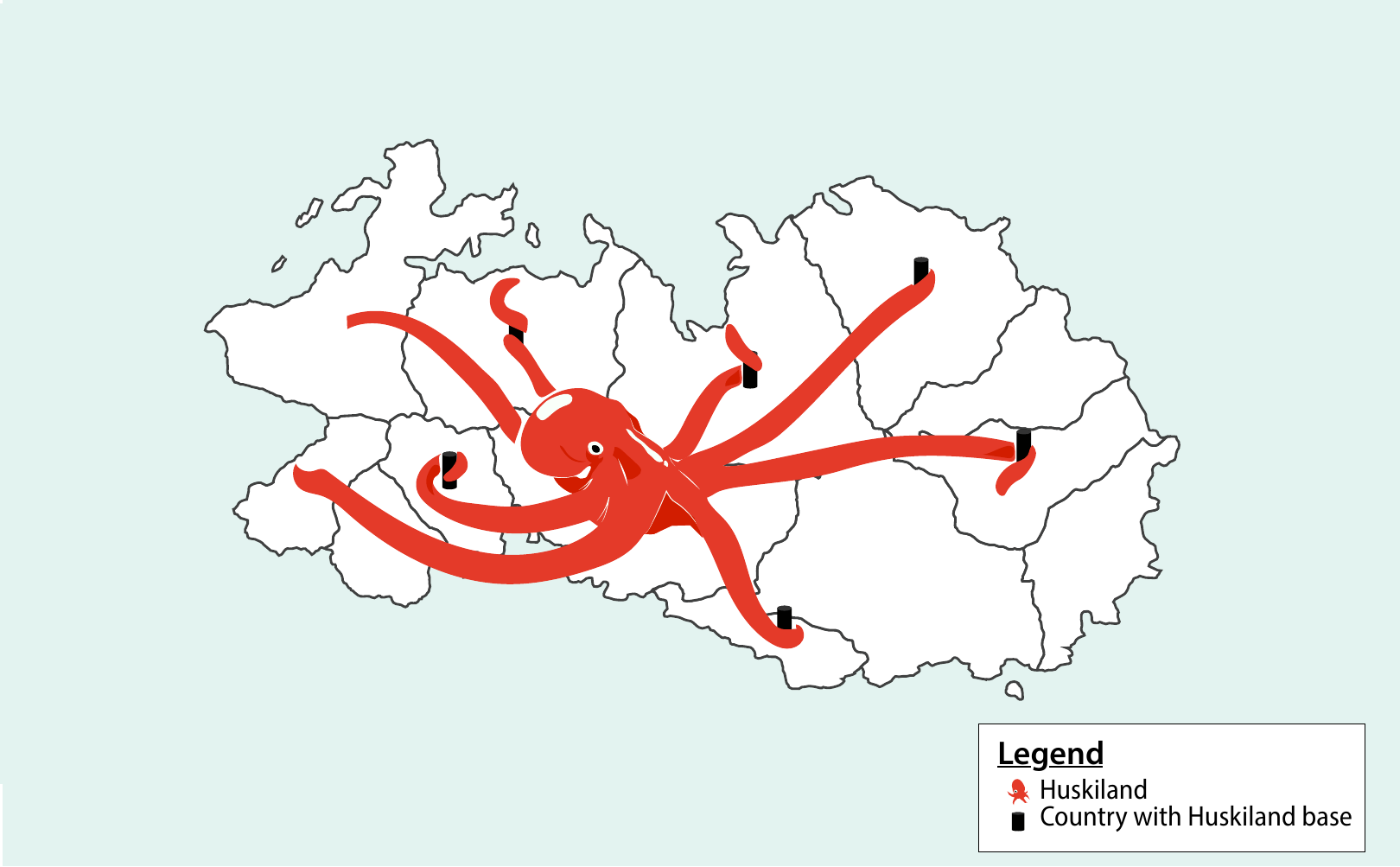}} & 
    \adjustbox{valign=m}{\includegraphics[width=0.22\textwidth]{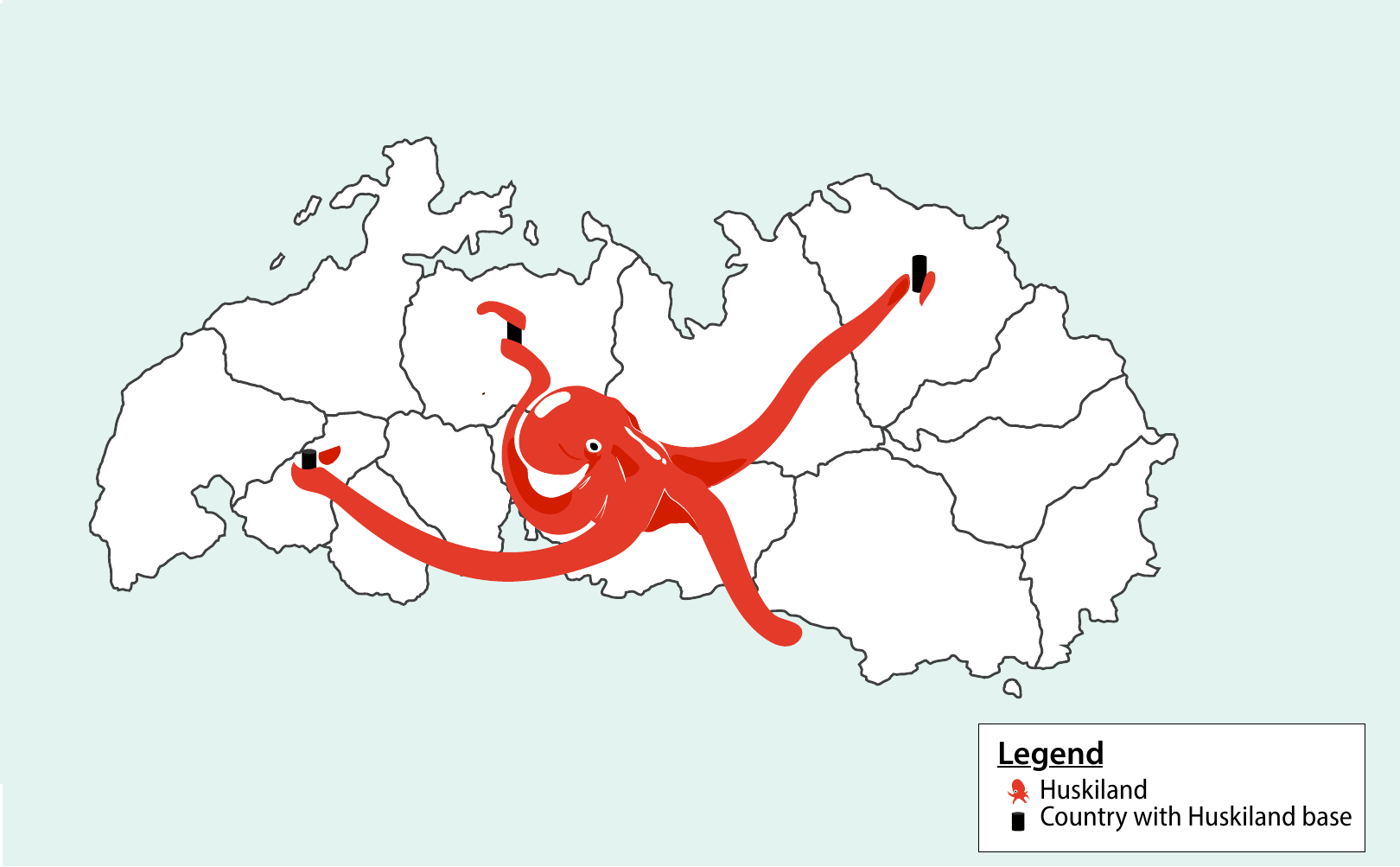}} \\
    
\end{tabular}
\end{center}
    
    \Description{Sixteen different maps of a fictional country ``Huskiland'' with varying numbers of countries around it highlighted or otherwise indicated as having Huskiland bases. These connections are indicated via color alone, via node-link lines, arrows, or via an icon of an octopus with tentacles in each country. These indications are either neutral gray or menacing red.}
    \caption{The sixteen stimuli from our crowdsourced experiment, showing the arrangement of extraterritorial military bases around a fictional country, Huskiland. We varied the \textbf{Color} (either a gray meant to be neutral, or a red meant to be more sensationalized), \textbf{Edge Type} (representing connections via color, node-link markers, arrows, or an octopus motif), and \textbf{Connection} (either four or eight external bases).}
    \label{fig:sample-stim}
\end{figure*}

Altogether, these factors resulted in 2 x 2 x 4 = 16 total maps. \autoref{fig:sample-stim} shows all of these combinations. Full size versions are included in our supplemental material.

\subsubsection{Participants}
We recruited our participants from the Prolific\footnote{\url{https://www.prolific.com/}} crowdworking platform, limiting our participant pool to residents of the United States who were between the ages of 18-65 and self-reported as fluent in English. We compensated participants $\$1$, based on a target rate of $\$12$/hour and an expected task duration of $5$ minutes as assessed from internal piloting. Metrics from Prolific reported an actual average task duration of $3$ minutes $30$ seconds, for an observed rate of $\$17.16$/hour.

We recruited a total of $300$ participants, of whom $44$ failed our preregistered exclusion criteria, for a final participant pool of $256$ participants. We note that, of those who failed the exclusion criteria (which was to correctly answer ``In how many countries in the region does Huskiland have military bases?''), $40/44$ ($91\%$) gave an answer that was only one number off, indicating a potential ambiguity in the question or stimuli rather than necessarily inattention. We include the full data set of $300$ participants in our supplemental materials but, as per our preregistration, present analyses of only $256$. The supplement also includes an analysis of the frequency of these validation failures by condition as an (admittedly incomplete) form of attrition analysis~\cite{zhou2016pitfall}, where---in line with prior work such as Moere et al.~\cite{moere2012style} that suggest that stylistic embellishments do not reliably impede the accuracy of reading data values---we did not observe any consistent pattern of errors among conditions, although we note that our study was not sufficiently powered nor were our methods for detecting attrition sufficiently sophisticated to detect the systematic patterns of attrition that might indicate failure in our assumption of random assignment.

Of the $256$ participants, $134$ were female, $115$ male, and $7$ reported other gender identities or declined to state. In terms of age, the plurality of participants ($95$) were between 25-34 years old. In terms of education, the plurality ($109$) had the highest educational attainment of a bachelor's degree. On a 7-point rating scale, participants generally reported a high familiarity with graphs ($M=6.0$, $SD=0.9$), maps ($M=6.0$, $SD=1.1$), and, to a lesser extent, political cartoons ($M=5.1$, $SD=1.5$). Additional demographic information is reported in our supplemental material.

\subsubsection{Analyses}
We preregistered two sets of analyses. The first was a quantitative analysis of \textit{aggregate} rating scores with respect to the map condition, our summative scale of ``octopodality'', which we intended to function as the extent to which our participants agreed with what we view of as the overall visual argument of the canonical octopus map. We summed all six 1-7 rating scales into a single metric that corresponds with the degree of agreement with an overall octopus-like visual argument (as laid out in \autoref{sec:parts}). In keeping with other related scales (such as propensity to conspiratorial thought~\cite{brotherton2013measuring}, octopodality was a simple summation of items, resulting in valid values from 6 (if a participant rated all items as ``1'') to 42 (if a participant rated all items as ``7''). We preregistered a one-way ANOVA on the impact of the 16 map conditions on overall rating, using Tukey's Test of Honest Significant Difference (HSD) as a post-hoc test to generate clusters of maps with similar ratings. While we did not have any specific pairwise map hypotheses, we did predict (but did not instantiate into specific hypotheses) that maps with more sensationalized ``octopus-like'' components, either identified by us as hallmarks of octopus maps or by others as hallmarks of sensationalist maps (i.e., threatening colors or arrows) would cluster together with higher ratings.

The second set of preregistered analyses investigates the connection between our aggregate ``octopodality'' rating and the overall sentiment derived from the visual argument of the map. We note that our prompt was free of explicit information on the role of Huskiland with respect to its neighbors, or the valence of the relationship, and that, even in the sensationalized octopus condition, there are no markers (like frowning faces, dripping blood, knives or other weapons) that have been used in other octopus or octopus-like maps as tools to reinforce the malevolent intent of the octopus. We estimated sentiment, taking the response to our free-text question and coding it as either \textit{Positive} (indicating that Huskiland was seen as a benevolent, protective, or otherwise positive force), \textit{Negative} (Huskiland as a threatening, controlling, or otherwise negative force), and \textit{Neutral} (mentioning both positive and negative aspects of the relationship, or where the response was too ambiguous to assign polarity). We opted for manual coding of trinary sentiment analysis rather than the use of automated sentiment analysis methods for the following reasons: one, since each participant was providing short textual responses to a single map, we did not think the resulting responses would be long or detailed enough to reliably produce meaningful or consistent automated sentiment scores. Two, our decision to intentionally restrict the size and complexity of the experimental meant that we were able to read each response in detail, which we believed to afford more detailed and accurate interpretations than automated sentiment analysis methods. Reading the responses in detail allowed us to conduct some additional qualitative analysis, which we discuss in \autoref{subsec: qualitative}. Three, we were interested in the specific rhetorical impact of the stimuli and reviewing participants' own descriptions of the relationship between Huskiland and its neighbors allowed for a deeper understanding of their interpretations (investigating which parts of the implicit argument appeared to have the most impact, rather than a shift of aggregate polarity) complementing the qualitative and quantitative analysis conducted on the responses. In the supplemental material, we include a comparison of our manual tags with the Qualtric's ``TextIQ''\footnote{https://www.qualtrics.com/support/survey-platform/data-and-analysis-module/text-iq/sentiment-analysis/} sentiment analysis model, highlighting areas of substantial (but important) disagreement.

Our hypothesis was that participants who responded higher on our octopodality scale, and so imbued the map with more of the components we felt were intrinsic to octopus-like visual arguments, would be more likely to describe Huskiland negatively, as detected in a one-way ANOVA of the impact derived sentiment on aggregate rating, with a post-hoc Tukey HSD test to determine which, if any, of the three sentiment groups were significantly different.

\textit{Deviations from preregistration}: we made two notable deviations from our preregistration. The first is that, while we had initial predictions about the relative rankings of maps with respect to our aggregate measure, we did not have strong hypotheses about the relative impact of individual factors (which we viewed merely as tools for generating plausible maps rather than an a priori structured or complex design space) and so planned on an analysis treating each of the 16 map conditions independently. However, post-hoc analysis showed a ranking where maps with low \textbf{Connection} were consistently rated lower on our rating scales than maps high \textbf{Connection}. To explore these sorts of patterns we performed a second, three-way ANOVA of the impact of \textit{Color}, \textit{Edge Type}, and \textit{Connection} on aggregate rating. We caution that this analysis was \textit{post-hoc}, and that the experimental design is underpowered with respect to assessing three-way interactions.

The second deviation concerns the qualitative coding: we had preregistered two coders would independently code results, and then would meet to resolve any disagreements. However, all three authors had time to independently code, and so we opted for more coders. Of the mismatches ($59/256=23\%$ of codes), we resolved differences with voting ($57/59$ of mismatches), or additional discussion if there was still a disagreement in polarity ($2/59$ of mismatches). The use of three coders meant that we could not use our preregistered metric of inter-rater reliability, Cohen's $\kappa$, which is only appropriate for a single  pair of raters. We instead opted for the conceptually similar Fleiss' $\kappa$, which was $0.764$, or ``substantial agreement'' as per the categorical levels of Landis \& Koch~\cite{landis1977measurement}.

Our preregistration includes several areas of analysis (such as exploration of emergent patterns for qualitative responses), where we had no strong hypotheses but wished to reactively explore any patterns of interest. Since these analyses were by nature \textit{post hoc}, as per our preregistration, we explicitly label them as exploratory, and report them separately in \S\ref{sec:exploratory}.

Lastly, our supplemental material includes additional \textit{post hoc} analyses around scale construction, correlation, and attrition analysis that were not part of our preregistration (nor connected to our preregistered hypotheses), but that provide additional details that contextualize our results or address methodological concerns.

\subsection{Preregistered Results}

\begin{figure*}
    \centering
    \includegraphics[width=\textwidth]{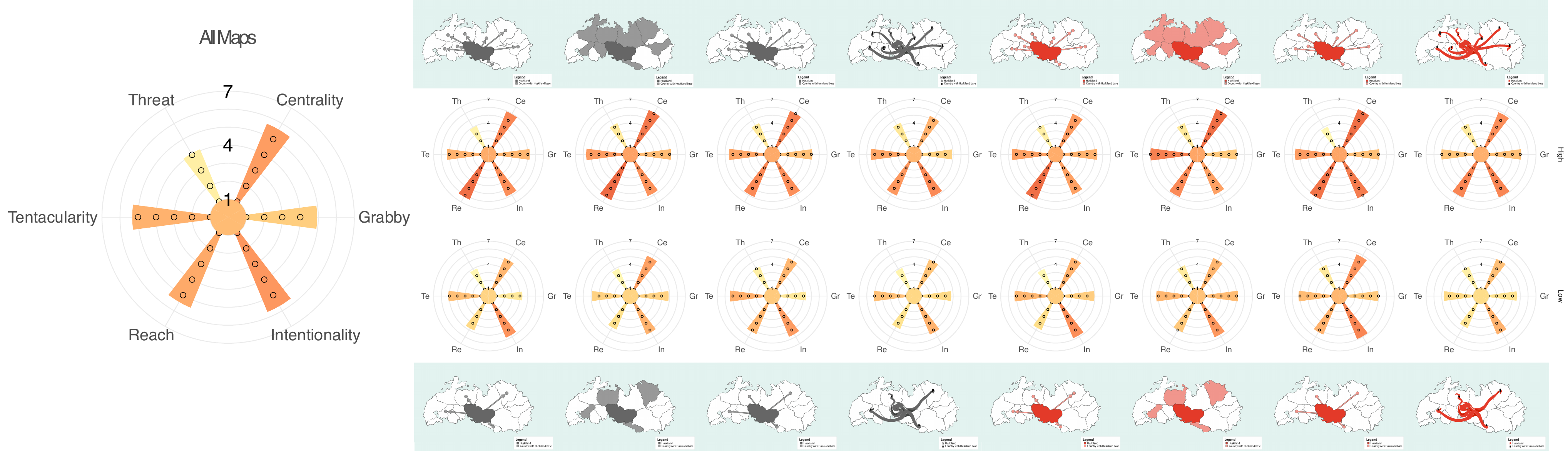}
    \Description{A set of star glyph charts presenting the results of our study for each type of stimuli.}
    \caption{Star glyphs of our results for each map condition shown to participants, split up by maps with high (top row) and low (bottom row) \textbf{Connection}. Each arm redundantly encodes participant's average Likert scale ratings along our six axes of octopodality as both hue and length, and the center of each glyph is colored based on the average across all six axes. The large glyph on the left shows the average rating for each of our sixth octopus aspects across all conditions. There were no significant pairwise difference in aggregate rating across map types, but there were significant differences between maps with high or low connection.}
    \label{fig:allMaps}
\end{figure*}

\begin{figure}
    \centering
    \includegraphics[width=0.5\textwidth]{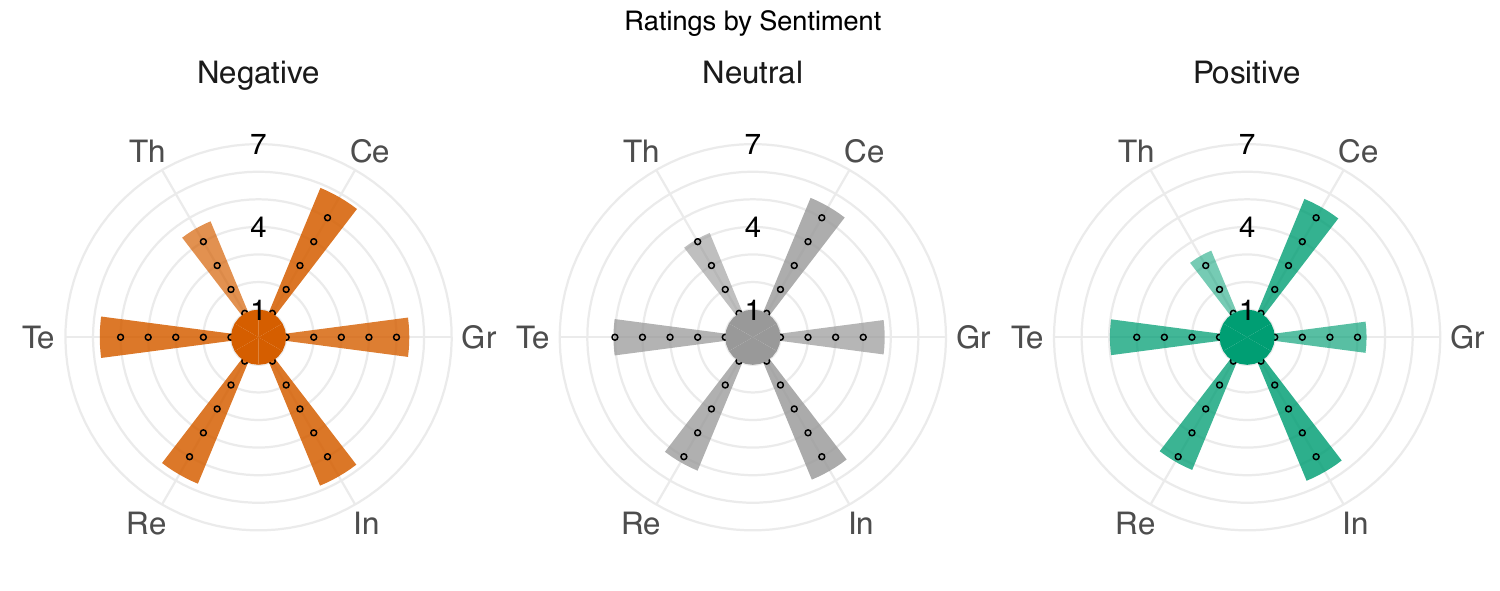}
    \Description{A set of three star glyphs, each presenting the results respectively associated with responses the authors coded as negative, neutral, and positive.}
    \caption{Star glyphs of the relationship between the sentiment polarity we coded in participant's free text responses, and the participant's  Likert scale rating of the maps they saw along our six axes of octopodality. Participants who described the relationship between Huskiland and its neighbors in negative terms rated the maps significantly higher on average than those who were positive, neutral, or whose stance we could not determine.}
    \label{fig:sentiment}
    \vspace{-1em}
\end{figure}

All analyses and figures are available in our supplemental material as an R markdown file. In keeping with the theme of the paper, we graphically report our results using octopus-like star glyphs~\cite{van2022out}. More traditional charts are available in the supplemental material.

A one-way ANOVA indicated that map type was a significant factor in average aggregate rating ($F(15,240)= 2.82$, $p=0.00046$). However, a post-hoc Tukey's HSD did not identify any statistically significant pairwise differences.

The modal participant described the relationship between Huskiland and its neighbors in terms we judged to be negative ($106/256 = 41.4\%$), although the plurality of participants described the relationship in either neutral ($77/256 = 30.1\%$) or positive ($73/256 = 28.5\%$) terms. There was a connection between the sentiment of the free-text used to describe the relationship between Huskiland and its neighbors, and aggregate rating. A one-way ANOVA of the impact of sentiment on aggregate rating found a significant difference ($F(2,253)=26.9$, $p=2.55e-11$). A post-hoc Tukey's HSD found that participants who described the relationship negatively rated the maps significantly higher in aggregate rating ($M=33.2$, $SD=3.8$) than those who used neutral ($M=30.2$, $SD=4.0$) or positive ($M=28.9$, $SD=4.2$) language. \autoref{fig:sentiment} explores this result in additional detail. We note, however, that we do not propose, nor does our experimental design support, a particular causal linkage underlying this effect (e.g., if those who view the relationship as negative are more likely to rate the map highly, or vice versa).

\subsection{Exploratory and Additional Analyses}
\label{sec:exploratory}

\subsubsection{Exploratory Quantitative Analyses}
In our analysis of aggregate octopodality above, we noticed a consistent clustering where maps with high \textbf{Connection} were rated more highly than maps with low \textbf{Connection}. \autoref{fig:allMaps} shows our results in more detail, broken out by all 16 map conditions and 6 rating items. In a deviation from our preregistration, we conducted a three-way ANOVA treating \textbf{Color}, \textbf{Edge Type}, and \textbf{Connection} as independent factors. \textbf{Connection} emerged as having a significant impact on overall rating ($F(15,240)=32.9$, $p=2.94e-08$). A post-hoc Tukey's HSD indicated a significant difference in aggregate rating between maps, with low \textbf{Connection} rated lower in aggregate ($M=29.6$, $SD=4.2$) than maps with high \textbf{Connection} ($M=32.6$, $SD=4.1$). Both the study authors and some participants remarked on the visual properties of maps with differing levels of \textbf{Connection}. In the low \textbf{Connection} condition, where the octopus has only four tentacles that touch only immediate neighbors, one participant (P8) even described the octopus as ``\textit{cute}.'' (c.f. P280 in the high \textbf{Connection} condition, where Huskiland's neighbors were ``\textit{wrapped in tentacles that don't seem to want to let go.}'').

While we had no strong hypotheses as to the degree to which the six components of the visual argument of the octopus map would be reported by participants, and focused our analyses on the aggregate of these features to better capture the visual argument as a whole, we did observe variability within responses, as visible in the aggregate glyph on the left of \autoref{fig:allMaps}. We performed a one-way ANOVA on the impact of each of our six scale items on individual rating and found a significant effect ($F(5,1530)=75.4$, $p=<2e-16$). A post-hoc Tukey's HSD identified that our question around \textbf{Threat} was rated significantly lower on average ($M=4.08$, $SD=1.15$, compared to an average of $M=5.18$ for all ratings) than all other ratings, with our question around \textbf{Grabby}-ness rated significantly higher than \textbf{Threat} but lower than the remaining four items ($M=4.93$,$SD=1.18$). As mentioned, we had no strong hypotheses around potential differences in these ratings, but we do note that both the \textbf{Grabby} item and \textbf{Threat} include terms, ``military or political control'' and ``threat'', respectively, that are connected with negative attitudes in ways that are less prominent for the other items. The supplemental material includes a more detailed analysis of the correlation between our scale metrics.

Lastly, while our preregistered quantitative analyses focused on the \textit{differences} in octopodality among maps, we note the \textit{overall high ratings} on our scale ($M=31.1$. $SD=4.4$ out of a maximum potential value of $42$). The combination of a lack of significant difference between map types, an (admittedly exploratory) significant difference between a data-based (\textit{Connection}) rather than design-based map feature, and this generally high rating suggests that what we term to be the overall visual argument can be successfully conveyed even in maps that eschew what we view as overtly sensationalized visual metaphors and designs.

\subsubsection{Exploratory Qualitative Analysis}
\label{subsec: qualitative}

As mentioned, we intentionally chose a fictional map and scenario in order to minimize (although, admittedly, not eliminate) the impact of prior knowledge or conceptions in the rhetorical impact of our maps. Nevertheless, given the iconic connection of octopus maps and historical propaganda campaigns, we performed an exploratory analysis of the responses to our free-text question (``How would you describe the relationship between Huskiland and its neighbors in the region?'') both to observe any connections to real-world or historical scenarios as well as to assess the rationale behind judgments in polarity in finer detail. In keeping with our intentionally narrow scope of the experimental task (a short response to a single map), we note that these responses comprised on average less than 20 words (\textit{M = 16.35, SD = 12.42}): about 35\% of responses were 10 words of less (more information on the distribution of response lengths is in the supplemental material). As such, while all paper authors kept notes of emergent themes and interesting features, we did not perform a formal thematic analysis.

One relevant question was the extend to which our fictional setting was or was not demonstrably entangled with existing framing and prejudices around real-world geopolitics. One paper author conducted a review of the free text responses looking for real-world political allusions, and then an additional analysis of term frequency based on common terms encountered during our pre-registered qualitative analysis. The full results of these searches are available in our supplemental material.

In keeping with our U.S. participant pool, we found that some participants framed their responses through the lens of US-centered current or historical events and alliances, with some responses invoking the US and its allies in describing the relationship between Huskiland and its neighbors. In the cases of two such responses, Huskiland's relationship with its neighbors is described in positive or neutral terms. For example, P16 commented, ``my assumption is that Huskiland and its neighbors have a military defense treaty similar to the US and its allies, with the US having military bases in most of its allies countries.'' P87 found the relationship ``similar to the united states to the countries that they put military bases in.'' Still, P15 found the relationship to be context-dependent, similar to our discussion in \S\ref{sec:canon}, noting, ``It could be protective, like the NATO alliance or aggressive like the AXIS [\textit{sic}] alliance.''

Another emergent research question was around the particular characterization of the relationships (and the extent to which these words aligned with the ``graspy'' and ``tentacular'' components of the implicit argument underlying our stimuli), in particular across our categories of overall sentiment. One paper author also ran an analysis of tokens used in responses and found that, excluding stopwords, some of the same words were frequently used within responses coded as either positive or negative. Within the positively coded responses, some of the most commonly used words include variations of ''allies'' (used 13 times in responses coded as positive, 3 for negative, 2 for neutral), ``protect'' (used 11 times in responses coded as positive, 1 for negative, 5 for neutral), and ``friend'' (used 11 times in responses coded as positive, 2 for negative, 2 for neutral). Positive responses were also more likely to use variations of the words ``allow'' (used 9 times in responses coded as positive, 1 for negative, 2 for neutral) and ``agreement'' (used in 6 times in responses coded as positive, 0 for negative, 4 for neutral), characterizing the relationship between Huskiland and its neighbors as a willing partnership. Some of the most commonly used non-neutral words used in responses coded as negative include variations of ``control'' (used 26 times in responses coded as negative, 1 for positive, 2 for neutral) and ``power'' (used 19 times in responses coded as negative, 4 for positive, 6 for neutral). We also found that participants whose responses were coded as negative used some version of the words ``tense'' (9 for negative, 0 for positive and neutral), ``invade or invasive'' (4 for negative, 0 for positive and neutral), and ``oppress'' (4 for negative, 0 for positive and neutral) to describe the relationship between Huskiland and its neighbors.

\section{Discussion}

Octopus maps are a cohesive visual genre of persuasive cartography with well over a century of precedent. These maps have an intended visual argument in which an entity is cast as a central threatening body with wide reaching connections, acquisitive intent, and a nefarious goal. Beyond the literal use of an octopus, octopus maps have commonalities in the visual strategies they use to persuade and the structure of the data they use for this persuasion. Crucially, these maps can take on rhetorical frames~\cite{hullman2011rhetoric} that can implicitly promote ``octopus-like'' visual arguments even without literal cephalopod imagery. The implicit or explicit visual arguments that arise from octopus maps are useful for many potential rhetorical goals: to villainize an enemy in wartime, cast aspersions on ideologies viewed as subversive or dangerous, or to promote (potentially conspiratorial) connections between events that would otherwise seem unrelated. It is possible to identify ``octopus-like'' framings and design patterns in a wide range of maps and visualizations.

Our results suggest that the presence of a literal octopus is not necessary for viewers to assume negative intent or produce octopus-like conclusions, even with relatively little prompting. We find that even maps we designed to be relatively free of overtly sensationalized persuasive elements could still engender negative sentiments and attributions of ill-intent in similar magnitudes to designs more directly inspired by propaganda posters and conspiracy theorists. Furthermore, while we caution the uncritical use of this evidence, our exploratory analysis suggests that the structure of the data (i.e., the number and connection of nodes) may have a larger impact on the rhetorical success of a sensationalized map than more traditional flourishes, like threatening arrows or grasping tentacles. The very act of framing a country as having large numbers of unilateral (although, admittedly in our experimental framing, explicitly militaristic) connections to its neighbors may intrinsically suggest adversarial and acquisitive relationships.

However, we note that the selection and framing of these data is just as much a design choice as how edges and nodes are visualized. An example from Muehlenhaus~\cite{muehlenhaus2014going} shows how the large number of U.S. military assets near Iran can support two opposing octopus-like framings. Per one map, published in an article in the Daily Mail, ``How Iran can strike US targets in the Middle East: Missiles, sea mines, drones and battle-hardened jihadist militias stand ready throughout the region amid mounting tensions''~\cite{iranBad}, Iran is cast implicitly as an octopus-like threat with multiple tentacular avenues of attack throughout the region. Yet, an Al Jazeera infographic, ``Map: US bases encircle Iran''~\cite{usBad} uses similar data to implicitly depict an American octopus, with multiple nearby bases poised to attack internally into Iran.

We acknowledge that octopus maps are an extreme example of rhetorical structure in charts: the adversarial and conspiratorial nature of these maps are often obvious, relying on well-worn tropes and existing wellsprings of animus or conspiratorial thought. Still, other forms of data design are not free from such considerations. We charge designers to consider the implicit and explicit \textit{visual arguments} in their designs carefully and with intention. We also sign on to the underpinnings of projects like \textit{Data Feminism}~\cite{d2023data,criadoperez2019invisible} that hold that structures of oppression and exclusion make themselves felt in the ways that data are collected, structured, and visualized, sometimes without even conscious awareness of the designers of these visualizations.

\subsection{Design and Ethical Implications}

We recognize that, while the modal reader of this paper might encounter an octopus or octopus-like map in their day-to-day (a perhaps unfortunate side effect of this work is that the authors now have started seeing the hallmarks of these maps almost everywhere we look), they are unlikely to be specifically asked to design or evaluate an octopus map of their own. Our implications for design therefore concern---much as Rensink~\cite{rensink2018information} dubs scatterplots to be the ``fruit flies'' of graphical perception work for their utility as a test bed--- the extent to which octopus maps can serve as meaningful ``fruit flies'' for the space of visual rhetoric and persuasive cartography. While once again urging the reader to avoid over-interpreting or over-generalizing the results we produce from an idiosyncratic corpus and an intentionally tightly-scaled empirical analysis, we believe that our results can provide guidance (even if only inchoate) across three critical areas of visualization research, ethics, and design. 

\subsubsection{Sensationalization in Data Visualization and Cartography}

There is a temptation among visualization designers (especially in academic visualization) to (falsely) view one's work as ``the mere reporting or structuring of objective fact''~\cite{correll2019ethical}, with the work of persuasion or rhetorical appeals seen as the exclusive purview of bad or manipulative actors. Our results, however, suggest that cartography and visualization cannot be neatly divided into sensationalized ``persuasive'' maps and charts versus more ``neutral'' charts without overt rhetorical intent. Persuasive elements are not merely a visual style that is applied on top of otherwise neutral data to render it sensationalized; choices of how to frame, select, and structure the underlying data can all perform important rhetorical or persuasive work, regardless if edges in a graph are grasping red tentacles or neutral gray lines. Even designers who are careful to avoid the design hallmarks of propaganda maps can still find their audiences captured by conspiratorial or adversarial views.

Readers, too, bring their own expectations and contexts with them to maps and charts, further complicating the possibility of any clear dichotomy between neutral or sensationalized views of data. The good intent of visualization designers is not sufficient to avoid moral culpability for how visualizations can be misinterpreted or misused for conspiratorial ends. An example is the rhetorical use of visualizations of COVID-19 data~\cite{lee2021viral}, where seemingly laudatory data literacy activities like questioning sources, assessing positionality, and conducting alternative analyses were used to support conspiratorial thinking. 

We do, however, echo the advice of van Houtum~\cite{van2020migration} and the principles in resources like the ``Do No Harm Guide''~\cite{schwabish2021no} to avoid the (sometimes subtle) design choices that can marginalize or disparage vulnerable populations. While our empirical findings suggest that some of the impacts of avoiding such visual design choices may be small or stochastic, or at the very least dwarfed buy other considerations in how maps are interpreted, the duties and responsibilities to the people represented in our datasets charge us to take every effort to avoid the language of invasion and subversion, visual or otherwise. We also charge readers of data visualizations to avoid the temptation to be satisfied by a surface-level judgment that, just because a visualization lacks the hallmarks of a sensationalized map, that it is therefore free from the many ways that designers can manipulate or mislead their audiences~\cite{lisnic2023misleading,lisnic2024yeah}.

\subsubsection{Visual Metaphors}

The use of the metaphor of the octopus is entangled with a wide array of implicit and explicit meanings, some of which are situated in specific historical and cultural contexts. Attempting to understand the octopus \textit{map} without understanding the octopus \textit{motif} is to elide large portions of the designer's intent on how these maps are meant to be read. However, we hold that these sorts of entanglements are not unique to extreme examples like the octopus map, but that the complexities of visual metaphors run through even the most quotidian of charts and maps. For instance, as per Ziemkiewicz and Kosara~\cite{ziemkiewicz2009preconceptions}, whether nodes in a tree are perceived as being \textit{contained by} their parents or are \textit{descended from} their parents can be influenced by both the visual representation used and also the preconceptions of the viewer.

The impact of a visual metaphor is influenced by a number of factors (like familiarity, priming, and expressiveness), not all of which are directly amenable to standard evaluative methods in graphical perception. Exploring or evaluating a visualization for its \textit{metaphorical} content may look more like Bares et al.'s \textit{close reading}~\cite{bares2020close}, or Pokojná et al.'s \textit{deconstruction}~\cite{pokojna2024language} than a straightforward quantitative assessment of visual encodings with respect to their legibility or efficiency. More than suggesting the necessity of adapting or creating methods for visualization research or design, we believe that visual metaphors are also a crucial part of rethinking visualization pedagogy. Visualization literacy is potentially less of a process of learning how to decode or extract data values from charts but also a process of building familiarity with common visual metaphors and genres.

\subsubsection{Visualization Rhetoric}

As mentioned, at a low level of description of the underlying data, an octopus map is merely a juxtaposition of a graph structure with cartography. Yet, the emotional and propaganda purposes to which these maps have been employed suggest that there are intended rhetorical goals for these maps above and beyond simply reading and understanding the underlying data. There is more to the success or failure of such visualizations than the efficiency with which data values are extracted~\cite{bertini2020shouldn} (in Aristotlean terms, the \textit{logos} of a visual argument): the \textit{ethos} and \textit{pathos} of a chart are likewise increasingly critical to the chart's intended and actual impact~\cite{kostelnick2016re}. The octopus map is therefore a \textit{visual argument}~\cite{blair2012possibility}: it has an intended context, and intended audience, and intended interpretation.

Our empirical results that fail to identify an obvious difference between octopus maps and less overtly manipulative designs suggest one (or, likely, both) of two conclusions: either (as we suggest above) that this apparent boundary between propagandist and ``neutral'' cartographer is more porous than prior work assumed and/or that we \textit{lack the empirical tools to holistically assess the implicit rhetorics of data visualization}. While prior work has asserted that implicit rhetorics shape our conception of data visualization in ways large and small~\cite{kennedy2016work,kostelnick2003shaping}, empirical evaluation of these claims is likely to require new approaches and new methods. For instance, more precise elicitation methods~\cite{hullman2016evaluating}, better accounting for Bayesian priors~\cite{kim2019bayesian}, and longitudinal assessments of not just comprehension of a single data visualization, but of how the persuasive power of a data visualization integrates into a holistic information ecosystem of existing narratives and perspectives.

\subsection{Limitations}
We reiterate several key limitations in both of our historical analysis and crowd-sourced study.

For our historical analysis, we note that biases in what propaganda maps are or are not retained and archived likely prevents anything like a truly representative corpus of octopus maps. Our analysis of the corpus we \textit{did} collect likewise focuses on the visual and topical forms of these maps: we are not historians, and the sheer breadth of time periods, languages, cultural settings, and political viewpoints we encountered in our search necessitate lacunae in our understandings of the finer details or broader impacts of these maps.

With respect to the design of our crowd-sourced study, we note that our desire to keep the scope of the experiment relatively small and the amount of data collected in line with what we as a research team could meaningfully and individually analyze precludes meaningful empirical data on several pertinent research questions. Beyond our previously stated admission that our choice of stimuli includes only a fraction of the potentially relevant design and measurement space, our study design does not afford analysis of pre- and post-exposure shifts in attitudes (as in common in other works on visualization rhetoric~\cite{pandey2014persuasive}), and is not sufficiently powered to robustly identify complex interaction effects. We also note that, while our choice of a scenario using a fictional country was made in an attempt to ameliorate potential biases from real-world geopolitics, we are under no illusions that any truly ``bias-free'' prompt exists, and the prior assumptions or ready-to-hand examples of military relationships could (and, for some participants where their responses made explicit mention of historical or contemporary examples, \textit{did}) influence how our maps were read. A potential avenue of future work would be to see just how innocuous a task setting can be while still producing conspiratorial interpretations (as an example, whether ``octopus-like'' maps of airline destinations or trade routes could be read as negatively monopolistic or expansionist), although we are cognizant of the fragility and unreliability of such framing effects in crowd-sourced studies~\cite{dimara2017narratives}.

With respect to the results from our study, our preregistered results generally fail to identify reliable effects, and much of our statistically significant results are based on non-preregistered or exploratory findings that we would encourage readers to take with a grain of salt, at least in the absence of follow-on confirmatory work. We wish to avoid over-claiming, and, in lieu of conducting a large number of (potentially underpowered) analyses with a high number of ``researcher degrees of freedom''~\cite{simmons2011false}, default to providing sufficient detail and data in the supplemental material for future replication or re-analysis.
However, as mentioned, we view our negative findings as interesting in their own right, as they suggest at the very least that octopus maps are not obviously and reliably distinct from their less-sensationalizes cousins along the rhetorical components we identified, and as an existence proof that participants can and do draw conspiratorial and ``octopus-like'' conclusions from a variety of comparatively anodyne map designs with relatively little prompting.

\subsection{Conclusion}

The octopus map is a particularly striking example of a sensationalized form of data visualization: a form of map that draws on an extended history of conspiratorial thinking to promote a particularly sinister view of an entity and its relationship to a geographic expanse. Nation-states, corporations, ideologies, and even abstract concepts have all been portrayed as a central controlling octopus. Yet, the implicit and explicit \textit{visual argument} of the octopus map is not unique, but is visible in more subtle guises in the ways that even well-meaning designers choose to convey geographic flows and networks. These implicit octopus forms have rhetorical power, impacting the ways that viewers interpret the data underlying maps and charts, bending them towards adversarially and conspiratorial framings. In short, we hold that the many sinister tendrils of octopus maps extend throughout the ways that data are conceived of, visualized, and interpreted.








\begin{acks}
Thank you to Laura Anna Garrison, Jane Adams, and Racquel Fygenson for their feedback on this paper. We also acknowledge funding support from Khoury College.
\end{acks}

\bibliographystyle{ACM-Reference-Format}
\bibliography{main}

\appendix

\section{Figure Credits}
\label{sec:figure_credits}

\begin{itemize}
\item \autoref{fig:subway}: 1899 political cartoon ``The Menace of the Hour'' by George Luks for \textit{The Verdict} magazine. Public domain from \href{https://commons.m.wikimedia.org/wiki/File:OctopusTheVerdict1899.jpg}{Wikimedia Commons}.

\item \autoref{fig:oil}: 1904 political cartoon ``Next'' by Udo Keppler for \textit{Puck} magazine. Public domain from \href{https://www.loc.gov/pictures/item/2001695241/}{The Library of Congress Prints and Photographs}.

\item \autoref{fig:russia}: 1877 ``Serio-Comic War Map for the Year 1877''  by Frederick W. Rose. Public domain from \href{https://commons.wikimedia.org/wiki/File:Serio-Comic_War_Map_For_The_Year_1877.jpg}{Wikimedia Commons}.

\item \autoref{fig:prussia}: 1916 or 1917 propaganda poster ``The Prussian Octopus'' by an unknown author. Public domain from \href{https://commons.wikimedia.org/wiki/File:Il_Calamaro_o%27_%E2%80%99Pesce_Diavolo%E2%80%99_Prussiano.jpg}{Wikimedia Commons}.

\item \autoref{fig:england}: Circa 1917 propaganda poster ``Freiheit der Meere'' [Freedom of the Seas] by an unknown author. Public domain from \href{https://digitalcollections.hoover.org/objects/14104/freiheit-der-meere-}{The Hoover Institution Library and Archives}.

\item \autoref{fig:hitler}: 1935 cover illustration of an issue of the magazine \textit{Fight: Against War and Fascism} by an unlisted author. Public domain from \href{https://dlib.nyu.edu/fawf/books/tamwag_fawf000020/#1}{The Tamiment Library and Robert F. Wagner Labor Archives}.

\item \autoref{fig:usa}: 1930 cover of the 1917 book \textit{Ante los B\'arbaros} [Before the Barbarians] by Jos\'e Mari\'a Vargas Vila. Public domain from \href{https://es.wikisource.org/wiki/Archivo:Portada-ante-los-barbaros-001.png}{Wikimedia Commons}.

\item \autoref{fig:landlordism}: 1909 political cartoon by W. B. Northrop titled “Landlordism Causes Unemployment.” © Cornell University – PJ Mode Collection of Persuasive Cartography and reproduced under Creative Commons Attribution-NonCommercial-ShareAlike 3.0 (CC BY-NC-SA 3.0) Unported License. \href{https://digital.library.cornell.edu/catalog/ss:19343701} {Cornell University – PJ Mode Collection of Persuasive Cartography.}

\item \autoref{fig:rothschild}: 1894 illustration ``The English Octopus - It Feeds on Nothing but Gold'' from the pamphlet \textit{Coin's Financial School} by William Hope Harvey. Public domain from \href{https://commons.wikimedia.org/wiki/File:Coin%27s_Financial_School_-_Denver_Road.png}{Wikimedia Commons}.

\item \autoref{fig:benevolent}: 1904 print by Kiyochika Kobayashi, ``The Japanese Octopus of Port Arthur.'' Public domain from \href{https://www.loc.gov/item/2009615031/}{The Library of Congress Prints and Photographs}.

\item \autoref{fig:rhodes}: 1892 political cartoon published in Punch magazine and illustrated by English cartoonist Edward Linley Sambourne. Public domain from \href{https://commons.wikimedia.org/wiki/File:Punch_Rhodes_Colossus.png}{Wikimedia Commons}.
\end{itemize}

\end{document}